\documentclass[journal]{IEEEtran}
\usepackage[T1]{fontenc}
\usepackage[utf8]{inputenc}
\usepackage{amssymb}
\usepackage[pdftex]{graphicx}
\usepackage{subcaption} 
\usepackage{amsmath}
\usepackage{amsthm}
\usepackage{mathtools}
\usepackage{array}
\usepackage[noadjust]{cite}
\usepackage{algorithm}
\usepackage{algpseudocode}
\usepackage{xcolor}
\usepackage{tikz}
\usepackage{pgfplots}
\usepackage{flushend}

\DeclareUnicodeCharacter{2212}{−}
\usepgfplotslibrary{groupplots,dateplot}
\usetikzlibrary{patterns,shapes.arrows}
\pgfplotsset{compat=newest}
\usepackage[hidelinks]{hyperref}
\usepackage[acronym]{glossaries}
\usepackage{dutchcal}
\pgfdeclarelayer{background}
\pgfdeclarelayer{foreground}
\pgfsetlayers{background,main,foreground}
\usepgfplotslibrary{fillbetween}
\usepgfplotslibrary{external}
\usetikzlibrary{spy,backgrounds}
\usepackage[binary-units=true]{siunitx}
\usepackage{balance}

\makeatletter
\let\MYcaption\@makecaption
\makeatother

\usepackage{subcaption}

\makeatletter
\let\@makecaption\MYcaption
\makeatother
\algnewcommand{\LineComment}[1]{\(\triangleright\) #1}

\newacronym{ACLR}{ACLR}{adjacent channel leakage ratio}
\newacronym{PAPR}{PAPR}{peak-to-average power ratio}
\newacronym{QAM}{QAM}{quadrature amplitude modulation}
\newacronym{NN}{NN}{neural network}
\newacronym{AWGN}{AWGN}{additive white Gaussian noise}
\newacronym{3GPP}{3GPP}{3rd Generation Partnership Project}
\newacronym{OFDM}{OFDM}{orthogonal frequency division multiplexing}
\newacronym{RRC}{RRC}{root-raised-cosine}
\newacronym{CP}{CP}{cyclic prefix}
\newacronym{PRT}{PRT}{pilot reduction tone}
\newacronym{LLR}{LLR}{log-likelihood ratio}
\newacronym{DFT}{DFT}{discrete Fourier transform}
\newacronym{BICM}{BICM}{bit-interleaved coded modulation}
\newacronym{LMMSE}{LMMSE}{linear minimum mean square error}
\newacronym{BER}{BER}{bit error rate}
\newacronym{BMD}{BMD}{bit-metric decoding}
\newacronym{ISI}{ISI}{inter-symbol interference}
\newacronym{KL}{KL}{Kullback–Leibler}
\newacronym{PSD}{PSD}{power spectral density}
\newacronym{iid}{i.i.d.\@}{independent and identically distributed}
\newacronym{SGD}{SGD}{stochastic gradient descent}
\newacronym{SNR}{SNR}{signal-to-noise ratio}
\newacronym{LOS}{LoS}{line-of-sight}
\newacronym{NLOS}{NLoS}{non-line-of-sight}
\newacronym{CCDF}{CCDF}{complementary cumulative distribution function}
\newacronym{PSK}{PSK}{phase-shift keying}
\newacronym{OMA}{OMA}{orthogonal multiple access}
\newacronym{NOMA}{NOMA}{non-orthogonal multiple access}
\newacronym{DC}{DC}{direct current}
\newacronym{5GNR}{5G NR}{5G New Radio}
\newacronym{OTFS}{OTFS}{orthogonal time frequency and space}
\newacronym{GFDM}{GFDM}{generalized frequency division multiplexing}

\renewcommand{\vec}[1]{\mathbf{#1}}
\newcommand{\vecs}[1]{\boldsymbol{#1}}

\newcommand{\rv}{\vec{r}}
\newcommand{\sv}{\vec{s}}

\newcommand{\xv}{\vec{x}}

\newcommand{\zv}{\vec{z}}

\newcommand{\gammav}{\vecs{\gamma}}

\newcommand{\thetav}{\vecs{\theta}}
\newcommand{\psiv}{\vecs{\psi}}

\newcommand{\Am}{\vec{A}}
\newcommand{\Bm}{\vec{B}}

\newcommand{\Em}{\vec{E}}

\newcommand{\Xm}{\vec{X}}

\newcommand{\Cc}{{\cal C}}

\newcommand{\Lc}{{\cal L}}

\newcommand{\Rc}{{\cal R}}
\newcommand{\Sc}{{\cal S}}

\newcommand{\Uc}{{\cal U}}

\newcommand{\CC}{\mathbb{C}}

\newcommand{\RR}{\mathbb{R}}

\newcommand{\ZZ}{\mathbb{Z}}

\newcommand{\htp}{^{\mathsf{H}}}
\newcommand{\tp}{^{\mathsf{T}}}

\newcommand{\LB}{\left(}
\newcommand{\RB}{\right)}
\newcommand{\LP}{\left\{}
\newcommand{\RP}{\right\}}
\newcommand{\LSB}{\left[}
\newcommand{\RSB}{\right]}

\renewcommand{\log}[1]{\mathop{\mathrm{log_2}}\LB #1\RB}

\newcommand{\EE}{{\mathbb{E}}}

\newcommand\abs[1]{\left| #1 \right|}

\begin{document}
\title{Waveform Learning for Next-Generation Wireless Communication Systems}

\IEEEoverridecommandlockouts 

\author{Fay\c{c}al Ait Aoudia, \textit{Member, IEEE,} and Jakob Hoydis, \textit{Senior Member, IEEE}%

\thanks{F. Ait Aoudia and J. Hoydis are with NVIDIA, 06906 Sophia Antipolis, France (faitaoudia@nvidia.com, jhoydis@nvidia.com). Work carried out while both authors were with Nokia Bell Labs, France.}

\thanks{Parts of this work have been published in~\cite{aitaoudia2021end}.}
}

\maketitle

\begin{abstract}
We propose a learning-based method for the joint design of  
a transmit and receive filter, the constellation geometry and associated bit labeling, as well as a \gls{NN}-based detector. The method maximizes an achievable information rate, while simultaneously satisfying constraints on the \gls{ACLR} and \gls{PAPR}.
This allows control of the tradeoff between spectral containment, peak power, and communication rate.
Evaluation on an \gls{AWGN} channel shows significant reduction of \gls{ACLR} and \gls{PAPR} compared to a conventional baseline relying on \gls{QAM} and \gls{RRC}, without significant loss of information rate.
When considering a \gls{3GPP} multipath channel, the learned waveform and neural receiver enable competitive or higher rates than an \gls{OFDM} baseline, while reducing the \gls{ACLR} by \SI{10}{\decibel} and the \gls{PAPR} by \SI{2}{\decibel}.
The proposed method incurs no additional complexity on the transmitter side and might be an attractive tool for waveform design of beyond-5G systems. 

\end{abstract}

\glsresetall

\section{Introduction}


\Gls{OFDM} is the dominant waveform in current wireless communication systems, such as 4G, 5G, and WiFi, because it allows for a very efficient hardware implementation as well as single-tap equalization.
However, its sensitivity to frequency dispersion, reduction of spectral efficiency due to the cyclic prefix and guard bands, as well as high \gls{PAPR} and \gls{ACLR} make it unsuitable for future radio systems which are supposed to operate at very high carrier frequencies and as energy-efficiently as possible.
For example, the low power amplifier efficiency in the sub-THz bands calls for low PAPR waveforms that are able to deal with high phase noise while meeting strict requirements on out-of-band emissions~\cite{9083807}.
Moreover, the increased range of applications foreseen for beyond-5G systems could lead to a much wider range of waveforms available in future standards to efficiently fit the numerous edge cases, such as massive access from IoT devices in private industrial networks.
Many waveforms have been introduced to address some of the drawbacks of \gls{OFDM}, such as \gls{OTFS}~\cite{7925924} and \gls{GFDM}~\cite{5073571}. Also signal processing techniques on-top-of 
\gls{OFDM} were proposed to reduce its \gls{PAPR} and \gls{ACLR}, such as the use of guard subcarriers to improve spectral containment, and the allocation of \glspl{PRT} to reduce power peaks~\cite{1261335, 6633030}.
However, they require significant additional complexity on the transmitter side, and incur a rate loss as the subcarriers used as guards or \glspl{PRT} do not carry data.

Recent work has demonstrated the competitiveness of neural receivers for conventional waveforms~\cite{8052521, 8999801,9345504}.
However, such receivers can be trained to demodulate in principle \emph{any} waveform carrying information. Thus, they open the door to novel waveforms which cannot be easily demodulated by conventional techniques.
Building on this idea, we propose in this work to use end-to-end learning to jointly design a transmit and receive filter, a constellation geometry and the corresponding bit labeling, as well as a neural receiver.
The key idea of end-to-end learning~\cite{8054694} is to implement the transmitter, channel, and receiver as a \gls{NN}, that is trained to maximize an achievable information rate.
On the transmitter side, the widely-used \gls{BICM} architecture~\cite{669123} is preserved, such that no additional complexity is introduced compared to conventional systems. This has the additional benefit that an outer error-correcting code can be used. On the receiver side, the neural receiver computes \glspl{LLR} for the transmitted bits directly from the received samples. Optimization of the end-to-end system is performed to maximize the practical \gls{BMD} rate with constraints on the \gls{ACLR} and \gls{PAPR} of the waveform used for transmission.

The end-to-end learning approach was evaluated on both an \gls{AWGN} and a \gls{3GPP} multipath channel.
Results show that the proposed approach enables fine control of the tradeoff between the achievable information rate, \gls{ACLR}, and \gls{PAPR}.
On an \gls{AWGN} channel, a single-carrier waveform using \gls{QAM}, \gls{RRC} filtering, and Blackman windowing was used as benchmark.
The learning-based approach enables up to \SI{30}{\decibel} lower \gls{ACLR} compared to the baseline, while achieving a similar \gls{PAPR} and competitive rate.
Significant reduction of the \gls{PAPR} can also be achieved, however at the cost of a rate loss.
A multiple access channel involving two interfering users transmitting to a single receiver was also considered: two users share the same band, but each has their own trainable constellation and transmit filter.
The learned system enables rates greater than the ones achieved with \gls{QAM}, a \gls{BMD} receiver, and \gls{OMA}, by learning distinguishable constellations and transmit filters.
When considering a single-user \gls{3GPP} multipath channel, \gls{OFDM} with \gls{QAM} was used for benchmarking, with \gls{LMMSE} channel estimation and single-tap equalization.
Simulation results show that the learned waveform achieves similar or higher information rates than the considered baseline, while enabling \SI{10}{\decibel} lower \glspl{ACLR}, and \SI{2}{\decibel} lower \glspl{PAPR}.
To the best of our knowledge, this work is the first to propose a method for learning of constrained waveforms that achieve competitive performance.

The rest of this paper is organized as follows.
Section~\ref{sec:rw} presents the related work.
Section~\ref{sec:sm} introduces the system model, and Section~\ref{sec:pf} formulates the problem we aim to solve.
Section~\ref{sec:e2e} presents the end-to-end learning approach, and Section~\ref{sec:sr} the results of our simulations.
Finally, Section~\ref{sec:cl} concludes the paper.

\paragraph*{Notations}
Boldface upper-case (lower-case) letters denote matrices (column vectors);
regular lower-case letters denote scalars.
$\RR$ ($\CC$) is the set of real (complex) numbers; 
$()^*$ is the complex conjugate operator.
$\log{\cdot}$ denotes the binary logarithm.
The $(i,k)$ element of a matrix $\Xm$ is denoted by $X_{i,k}$. The $k^{th}$ element of a vector $\xv$ is $x_k$.
The operators $()\htp$ and $()\tp$ denote the Hermitian transpose and transpose, respectively. 
Finally, $j=\sqrt{-1}$ denotes the imaginary unit.

\section{related work}
\label{sec:rw}

The drawbacks of \gls{OFDM} have led to the introduction of many alternatives such as \gls{OTFS}~\cite{7925924}, whose key idea is to perform modulation in the delay-Doppler domain instead of the time-frequency domain, making it more robust to frequency dispersion.
Another example is \gls{GFDM}~\cite{5073571} which turns the channel into multiple parallel, independent, and possibly interfering single-carrier links.
Other examples of waveform proposals are available in~\cite{9390169} and references therein.
Concerning \gls{OFDM}, its high \gls{PAPR} and \gls{ACLR} are not novel concerns, see, e.g., \cite{6633030}.
In~\cite{1261335}, the authors propose to reserve subcarriers to the transmission of peak-canceling signals (\glspl{PRT}).
With this method, \gls{PAPR} reduction can only be achieved at the cost of a rate loss, as the subcarriers allocated to \glspl{PRT} do not carry data.
Moreover, the optimal \glspl{PRT} depend on the transmitted baseband symbols. \textcolor{black}{This implies} that an iterative and computationally demanding algorithm needs to be run for each \gls{OFDM} symbol, making this approach unpractical.
In~\cite{8928103}, the authors have extended the tone reduction method by unfolding a conventional algorithm to an \gls{NN}, which is trained offline to predict the \glspl{PRT} from the data symbols.
A similar approach was followed in~\cite{9419069}.
The authors of \cite{9037067} use a non model-based approach to predict the \glspl{PRT} from the data symbols with the help of an \gls{NN}. Still focusing on \gls{PAPR} reduction of \gls{OFDM}, it is proposed in~\cite{1233010} to extend the outer points of \gls{QAM} constellations with an active extension method.
As for \glspl{PRT}, the computation of the constellation extension needs to be carried out by an iterative and computationally demanding algorithm for each \gls{OFDM} symbol.
To address this issue, \cite{8928056} proposes to use an \gls{NN} to compute the constellation extension from the data symbols.

A different line of works consist in learning possibly high-dimensional modulations that operate on top of existing waveforms, typically \gls{OFDM}.
This approach was applied, e.g., in~\cite{aitaoudia2020,aitaoudia2021} to achieve pilotless and \gls{CP}-less communication, or in~\cite{9271932} to improve the \gls{BER}.
However, these works do not consider the \gls{PAPR} or \gls{ACLR}.
In~\cite{8240644}, the authors use an \gls{NN} to perform high dimensional modulation and demodulation on top of \gls{OFDM}, with the aim to improve the \gls{PAPR} while achieving low \glspl{BER}.
However, only the non-oversampled transmitted signal is used for the \gls{PAPR} computation, which is not completely representative of the analog waveform that actually serves as input to the power amplifier. Moreover, the \gls{ACLR} is not taken into account.
More recently, end-to-end learning of a high dimensional modulation and demodulation scheme operating on top of \gls{OFDM} was proposed in~\cite{goutay21}. Here, the transmitter and receiver were jointly optimized to maximize an achievable rate under predefined constraints on the \gls{PAPR} and \gls{ACLR}.
In the field of optical communications, end-to-end learning was used to improve the \gls{PAPR}~\cite{8971873,app9050852} and \gls{ACLR}~\cite{song2021end} of existing waveforms.

All of the references cited above propose methods operating on top of existing waveforms, such as \gls{OFDM}.
In this work, we present a method to design a novel single-carrier waveform by jointly optimizing the transmit filter and constellation geometry to maximize an achievable information rate while satisfying constraints on the \gls{PAPR} and \gls{ACLR}.
Compared to previous work, our approach does not require an \gls{NN} at the transmitter side and preserves the architecture and complexity of a conventional single-carrier system.
Moreover, by suitably designing the neural receiver, higher rates than those achieved by \gls{OFDM} are obtained on multipath channels, while preserving the benefits of the learned single-carrier waveform, i.e., low \gls{ACLR} and \gls{PAPR}.

\section{System model}
\label{sec:sm}

\begin{figure*}[h]
\centering{
\small
\def\svgwidth{\linewidth}
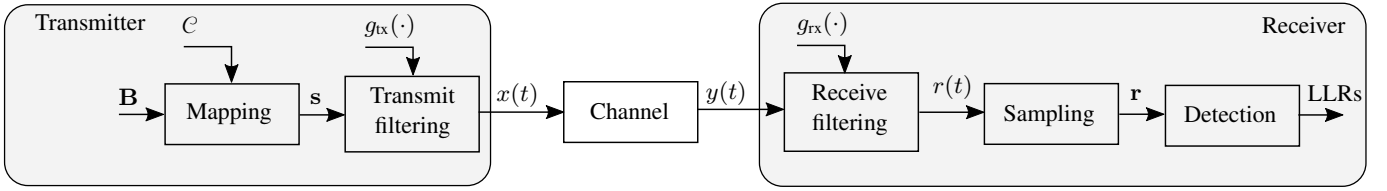
\caption{Architecture of a single-carrier system, where $\Cc$ is the constellation used for mapping, $g_\text{\text{tx}}(\cdot)$ the \textcolor{black}{transmit} filter, and $g_\text{\text{rx}}(\cdot)$ the receive filter.}
\label{fig:base_sys}
}
\end{figure*}

Let us denote by $\Bm \in \{0,1\}^{N \times K}$ the matrix of bits in a transmit block which consists of $N$ baseband symbols carrying each $K$ bits.
We consider \gls{BICM}, where $\Bm$ is modulated onto a vector of baseband symbols $\sv \in \Cc^N$ according to a constellation $\Cc \in \CC^{2^K}$ and a corresponding bit labeling, e.g., a $2^K$--\gls{QAM} with Gray labeling, or a learned constellation geometry and labeling.
We assume a classic single-carrier scheme in which the modulated symbols are shaped with a transmit filter $g_\text{\text{tx}} : t \in \RR \rightarrow \CC$ to form the time-continuous baseband signal
\begin{equation}
	\label{eq:x}
	x(t) = \sum_{n=0}^{N-1} s_n g_\text{\text{tx}}\LB t-n T \RB
\end{equation}
where $T$ is the symbol period.
This process is illustrated in Fig.~\ref{fig:base_sys}.

We consider a time-invariant multipath channel with baseband-equivalent transfer function
\begin{equation}
	\label{eq:h}
	h(\tau) = \sum_{p=0}^{P-1} a_p \delta\LB \tau - \tau_p \RB
\end{equation}
where $\delta\LB \cdot \RB$ is the Dirac delta function, $P$ is the number of paths, and $a_p$ and $\tau_p$ are the complex baseband channel coefficient and delay of the $p^{th}$ path, respectively.
The time-continuous channel output is given by
\begin{align}
	\label{eq:y}
	y(t) &= \int h(\tau) x(t - \tau) d\tau + w(t)\\
		 &= \sum_{n=0}^{N-1} s_n \sum_{p=0}^{P-1} a_p g_\text{\text{tx}}\LB t - \tau_p - n T \RB + w(t)
\end{align}
where $w(t)$ is uncorrelated Gaussian noise with power spectral density $N_0$, satisfying the condition $\EE \LSB w(t)w(t+\tau)^* \RSB = N_0 \delta(\tau)$.
\textcolor{black}{The noise is not assumed to be bandlimited.}

At the receiver, $y(t)$ is first filtered using a receive filter $g_\text{\text{rx}} : t \in \RR \rightarrow \CC$, leading to
\begin{align}
r(t)&= \int y(z) g_\text{\text{rx}}^*(t-z)dz\\
	&= \sum_{n=0}^{N-1} s_n \sum_{p=0}^{P-1} a_p \int g_\text{\text{tx}}(z') g_\text{\text{rx}}^* \LB t - z' - \tau_p -n T \RB dz' \nonumber\\
	&\quad+ \int w(z) g_\text{\text{rx}}(t-z)dz 
\end{align}
where the second equality comes from introducing $z' \coloneqq z - \tau_p - n T$.
The filtered received signal is then sampled with a period $T$ to form the vector $\rv \in \CC^N$ with elements 
\begin{equation}
\label{eq:trans_func_disc}
r_m = r(mT) = \sum_\ell s_{m-\ell} h_\ell + w_m
\end{equation}
where $\ell \coloneqq m-n$ and the channel taps $h_\ell$ are given by
\begin{equation}
\label{eq:ch_taps}
h_\ell = \sum_{p=0}^{P-1} a_p \LB g_\text{\text{tx}} \ast g_\text{\text{rx}}^* \RB (\ell T - \tau_p)
\end{equation}
where $\ast$ denotes the convolution operation and \textcolor{black}{$w_m$} is additive Gaussian noise with correlation
\begin{equation}
	\label{eq:noise_cov}
	\EE \LSB w_m w_{m+\ell}^* \RSB = N_0 \int g_\text{\text{rx}}^*(t) g_\text{\text{rx}}\LB t+\ell T \RB dt.
\end{equation}
As illustrated in Fig.~\ref{fig:base_sys}, $\rv$ is processed by a detector which computes \glspl{LLR} on the transmitted bits, which could then be fed to an outer decoder.

\subsection*{QAM with RRC filtering}
\label{sec:bsl}

As a baseline for comparison on the \gls{AWGN} channel, we consider the widely used \gls{QAM} with Gray labeling and \gls{RRC} filtering.
The \gls{RRC} filter is denoted by $\text{rrc}_{\beta}(t)$, where $\beta \in (0,1)$ is the roll-off factor which controls the tradeoff between the excess bandwidth and the magnitude of the ripples.
When used with its match filter $\text{rrc}_{\beta}(-t) = \text{rrc}_{\beta}(t)$, the \gls{RRC} filter satisfies the Nyquist \gls{ISI} criterion~\cite[Chapter~9.2.1]{proakis1994communication}
\begin{equation}
	\label{eq:nyquist_isi}
	\LB \text{rrc}_{\beta} \ast \text{rrc}_{\beta} \RB(\ell T) =
	\begin{cases}
		1 & \text{ if } \ell = 0\\
		0 & \text{ otherwise}
	\end{cases}
\end{equation}
where $\ell \in \ZZ$.
Because the \gls{RRC} filter is not time-limited, it is windowed by the Blackman windowing function $\text{bm}(t)$, which results in the following transmit and receive filters:
\begin{equation}
	g_\text{\text{tx}}(t) = g_\text{\text{rx}}(t) = \text{rrc}_{\beta}(t) \text{bm}\left(\frac{t}{D}\right)
\end{equation}
where $D > 0$ is the filter duration.

\section{Problem statement}
\label{sec:pf}

We will now present the optimization problem that we seek to address through end-to-end learning in Section~\ref{sec:e2e}.
As shown in Fig.~\ref{fig:base_sys}, we consider a single-carrier system, where the transmit and receive filters, denoted respectively by $g_{\text{tx},\thetav}(\cdot)$ and $g_{\text{rx},\psiv}(\cdot)$, have trainable parameters $\thetav$ and $\psiv$. Both filters are limited to the interval $\LB -\frac{D}{2}, \frac{D}{2} \RB$, i.e., for $t \notin \LB -\frac{D}{2}, \frac{D}{2} \RB$, $g_{\text{tx},\thetav}(t) = g_{\text{rx},\psiv}(t) = 0$.
On the transmitter side, the constellation geometry $\Cc$ and corresponding bit labeling are also trainable.
On the receiver side, the detector is implemented by a neural receiver with trainable parameters $\gammav$, which computes a posterior distribution $Q_{\gammav} \LB B_{n,k} | \rv \RB$, $0 \leq n \leq N-1$, $0 \leq k \leq K-1$, or equivalently \glspl{LLR}, on the transmitted bits $B_{n,k}$ from the received samples $\rv$.

Most practical systems rely on \gls{BICM} and \gls{BMD} for which the information rate
\begin{multline}
	\label{eq:R}
	R(\Cc,\thetav,\psiv,\gammav) \coloneqq \frac{1}{N} \sum_{n=0}^{N-1} \sum_{k=0}^{K-1} \Big[ I(B_{n,k};\rv | \Cc,\thetav,\psiv)\\
	- \EE_{\rv} \LSB \text{D}_{\text{KL}} \LB \Pr\LB B_{n,k} | \rv \RB || Q_{\gammav}\LB B_{n,k} | \rv \RB \RB \RSB \Big]
\end{multline}
is known to be an achievable rate~\cite{bmi}.
Therefore, we aim to maximize this rate.
Note that maximizing this rate typically leads to very different results than when maximizing the more conventional symbol-wise mutual information~\cite{9118963}, which is known to not be an achievable rate for practical \gls{BMD} receivers.
Moreover, using the \gls{BMD} rate~\eqref{eq:R} as objective enables joint optimization of the constellation geometry and bit labeling.
In~\eqref{eq:R}, $I(B_{n,k};\rv | \Cc,\thetav,\psiv)$ is the mutual information between $B_{n,k}$ and $\rv$ conditioned on the constellation and filters, $\text{D}_{\text{KL}}(\cdot||\cdot)$ is the \gls{KL} divergence, and $\Pr\LB B_{n,k} | \rv \RB$ is the \emph{true} posterior distribution on $B_{n,k}$ given $\rv$.
Intuitively, the first term on the right-hand side of~\eqref{eq:R} corresponds to an achievable rate assuming an optimal \gls{BMD} receiver, i.e., one that implements $\Pr\LB B_{n,k} | \rv \RB$.
The second term is the rate loss due to the use of a mismatched receiver, i.e., $Q_{\gammav}\LB B_{n,k} | \rv \RB \neq \Pr\LB B_{n,k} | \rv \RB$.

For practical purposes, the maximization of \eqref{eq:R} must be done under constraints on the transmitted waveform.
First, the transmit filter and constellation must have unit energy, i.e.,
\begin{align}
	&\int_{-\frac{D}{2}}^{\frac{D}{2}} \abs{g_{\text{tx},\thetav}(t)}^2 dt = 1 \label{eq:cst_en}\\
	&\EE_{c \sim \Uc\LB\Cc\RB} \LSB \abs{c}^2 \RSB = 1
\end{align}
where $\Uc\LB\Cc\RB$ is the uniform distribution on $\Cc$.

Concerning the \gls{PSD}, the \gls{ACLR} is defined as
\begin{equation}
	\label{eq:aclr}
	\text{ACLR}(\thetav) \coloneqq \frac{E_O(\thetav)}{E_I(\thetav)} = \frac{1}{E_I(\thetav)}-1
\end{equation}
where $E_I(\thetav)$ and $E_O(\thetav)$ are the in- and out-of-band energy, respectively, and the second equality comes from~\eqref{eq:cst_en} since $E_I(\thetav) + E_O(\thetav) = 1$.
As we consider \gls{BICM}, the sequence of modulated symbols $\sv$ is \gls{iid}, so that the in-band energy is
\begin{equation}
	\label{eq:inb}
	E_I(\thetav) = \int_{-\frac{W}{2}}^{\frac{W}{2}} \abs{\hat{g}_{\text{tx},\thetav}(f)}^2 df
\end{equation}
where $W \coloneqq \frac{1}{T}$ is the bandwidth and $\hat{g}_{\text{tx},\thetav}(f)$ the Fourier transform of $g_{\text{tx},\thetav}(t)$.

With regards to the power envelope, we denote by $p(t) \coloneqq \abs{x(t)}^2$ the signal power at time $t$.
Note that at every instant $t$, $p(t)$ is a random variable because of the randomness of the baseband symbols $\sv$.
Since the sequence of baseband symbols $\sv$ is \gls{iid}, $p(t)$ and $p \LB t + \ell T \RB$ share the same distribution for any $\ell \in \ZZ$ and $t$, such that $t$ and $t + \ell T$ are at least $\frac{D}{2}$ away from the temporal edges of the signal.
This can be seen from the definition of the transmitted signal~\eqref{eq:x}.
Therefore, it is sufficient to consider a single period $(-\frac{T}{2}, \frac{T}{2})$.
The \gls{PAPR} is defined as
\begin{subequations}
\begin{align}
\text{PAPR}\LB \thetav, \Cc \RB \coloneqq \quad &\underset{\nu \geq 0}{\text{minimize}}\quad \nu \label{eq:papr_def}\\
					\text{subject to\quad} \Pr & \LB \frac{p(t)}{\bar{p}} > \nu \RB = 0, t \in \LB -\frac{T}{2}, \frac{T}{2} \RB \label{eq:papr_ctr}
\end{align}
\end{subequations}
where
\begin{equation}
	\label{eq:avg_sig_pow}
	\bar{p} = \frac{N}{(N-1)T + D}
\end{equation}
is the average signal power.
\textcolor{black}{A derivation of~\eqref{eq:avg_sig_pow} is provided in the Appendix.}
\textcolor{black}{
One can notice that the equality~\eqref{eq:papr_ctr} is equivalent to
\begin{equation}
    \int_{-\frac{T}{2}}^{\frac{T}{2}} \max \LB \frac{p(t)}{\bar{p}} - \nu, 0 \RB dt = 0
\end{equation}
which is equivalent to
\begin{equation}
\EE \LSB \max \LB \frac{p(t)}{\bar{p}} - \nu, 0 \RB \RSB = 0 \label{eq:papr_equiv_ctr}
\end{equation}
where the expectation is over $\sv$ and $t \sim \Uc\LB -\frac{T}{2}, \frac{T}{2} \RB$.
}

Maximizing~\eqref{eq:R} with constraints on the \gls{ACLR} and \gls{PAPR} leads to the following optimization problem that we aim to solve
\begin{subequations}
\label{eq:prob_main}
\begin{align}
\underset{\Cc,\thetav,\psiv,\gammav}{\text{maximize}}\quad&
R(\Cc,\thetav,\psiv,\gammav) \label{eq:prob}\\
\text{subject to}\quad&
\int_{-\frac{D}{2}}^{\frac{D}{2}} \abs{g_{\text{tx},\thetav}(t)}^2 dt = 1 \label{eq:pc1}\\
&\EE_{c \sim \Uc\LB\Cc\RB} \LSB \abs{c}^2 \RSB = 1 \label{eq:pc2}\\
&\text{ACLR}(\thetav) \leq \epsilon_A \label{eq:pc3}\\
&\text{PAPR}(\thetav, \Cc) \leq \epsilon_P \label{eq:pc4}
\end{align}
\end{subequations}
where~\eqref{eq:pc1} and~\eqref{eq:pc2} constrain the energy of the transmit filter and constellation, respectively, and \eqref{eq:pc3} and~\eqref{eq:pc4} enforce the \gls{ACLR} and \gls{PAPR} to be less than or equal to $\epsilon_A$ and $\epsilon_P$, respectively.
From~\eqref{eq:papr_def},~\eqref{eq:papr_ctr}, and~\eqref{eq:papr_equiv_ctr}, one can see that the constraint~\eqref{eq:pc4} is equivalent to
\begin{equation}
V(\thetav, \Cc, \epsilon_P) \coloneqq \EE \LSB \max \LB \frac{p(t)}{\bar{p}} - \epsilon_P, 0 \RB \RSB = 0 \label{eq:pc4_equiv}
\end{equation}
where the expectation is over $\sv$ and $t \sim \Uc\LB -\frac{T}{2}, \frac{T}{2} \RB$.

\section{End-to-end waveform learning}
\label{sec:e2e}

\begin{figure*}[h]
\centering{
	\small
\def\svgwidth{\linewidth}
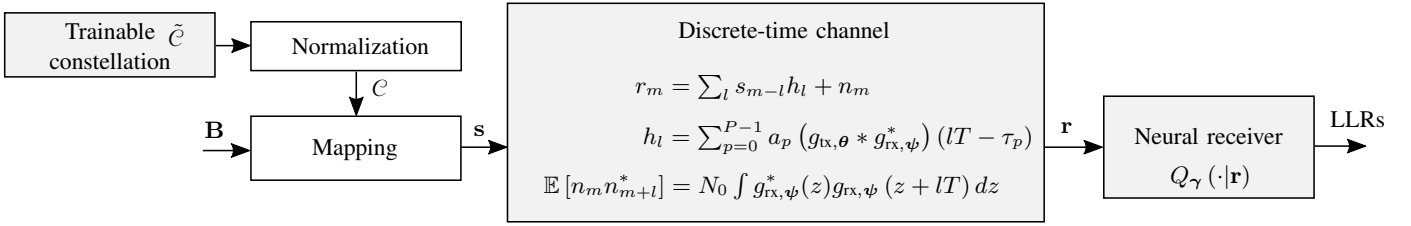
\caption{The trainable end-to-end system is implemented using a discrete-time channel model, which includes the effects of the transmit and receive filters. Components that include trainable parameters are shown in light gray.}
\label{fig:e2e}
}
\end{figure*}

End-to-end learning consists in implementing the transmitter, channel, and receiver as an \gls{NN} that is trained using \gls{SGD} by backpropagating the gradient from the receiver to the transmitter through the channel.
The considered end-to-end system is shown in Fig.~\ref{fig:e2e}, where the trainable components appear in light gray.
The channel is simulated in the discrete-time domain by implementing the transfer function~\eqref{eq:trans_func_disc}, which involves evaluating the channel taps~\eqref{eq:ch_taps} and noise correlation~\eqref{eq:noise_cov}.
The transmit and receive filters affect the received samples $\rv$ through the channel taps and noise correlation.
Training of the system through \gls{SGD} requires efficient computation of the channel taps and noise correlation.
These computations must also be differentiable.
This is made challenging as \eqref{eq:ch_taps} and \eqref{eq:noise_cov} involve integration.
The way how we solve this problem is detailed in Section~\ref{sec:tr_filters}.
Section~\ref{sec:const_tr} presents the implementation of the trainable constellation, and Section~\ref{sec:nn_tr} the neural receiver architecture.
Finally, in Section~\ref{sec:tr_proc}, we explain entire the training process.

\subsection{Trainable transmit and receive filters}
\label{sec:tr_filters}

As the computations of the channel taps~\eqref{eq:ch_taps} and noise correlation~\eqref{eq:noise_cov} require integration, we cannot easily use \glspl{NN} to implement the transmit and receiver filters as it would lead to intractable calculations. One could, e.g., approximate the integrals using Monte Carlo estimation or Riemann sums, but this is computationally demanding.
To accurately and efficiently implement trainable transmit and receive filters, we use the fact that the functions $\LP \text{sinc} \LB Df - s \RB \RP_{s \in \ZZ}$ form a frequency-domain basis of functions which are time-limited to $\LB -\frac{D}{2}, \frac{D}{2} \RB$.
We therefore define the trainable transmit and receive filters in the frequency domain as
\begin{align}
	\hat{g}_{\text{tx},\thetav}(f) &\coloneqq \sqrt{C(\thetav)} \sum_{s=-S}^{S} \theta_s \text{sinc} \LB Df - s \RB \label{eq:ftx}\\
	\hat{g}_{\text{rx},\psiv}(f) &\coloneqq \sum_{s=-S}^{S} \psi_s \text{sinc} \LB Df - s \RB \label{eq:frx}
\end{align}
where $S$ controls the number of trainable parameters.
The trainable parameters of the transmit and receiver filters are $\thetav = \LSB \theta_{-S},\dots,\theta_0,\dots,\theta_S \RSB\tp$ and  $\psiv = \LSB \psi_{-S},\dots,\psi_0,\dots,\psi_S \RSB\tp$, respectively.
Note that both filters are neither required to have the same number of parameters $2S+1$ nor the same duration $D$.
In~\eqref{eq:ftx}, $C(\thetav)$ is a normalization constant that ensures that the transmit filter has unit energy~\eqref{eq:pc1}.
Taking the inverse Fourier transform of~\eqref{eq:ftx} and~\eqref{eq:frx} leads to the time-domain expressions of the trainable filters
\begin{align}
	g_{\text{tx},\thetav}(t) &= \frac{\sqrt{C(\thetav)}}{D} \text{rect}\LB \frac{t}{D} \RB \sum_{s=-S}^{S} \theta_s e^{j2\pi\frac{s}{D}t} \label{eq:tx}\\
	g_{\text{rx},\psiv}(t) &= \frac{1}{D}\text{rect}\LB \frac{t}{D} \RB \sum_{s=-S}^{S} \psi_s e^{j2\pi\frac{s}{D}t}. \label{eq:rx}
\end{align}
One can see from these equations, that the transmit and receive filters are defined through Fourier series with $2S+1$ harmonics and period $D$, which are time-limited to a single period.

By using this filter parametrization, all of the quantities required to simulate the discrete-time channel have closed-form expressions that can be easily evaluated numerically:
\begin{align}
	C(\thetav) &= \frac{D}{\thetav\htp\thetav} \label{eq:c_norm}\\
	\LB g_{\text{tx},\thetav} \ast g_{\text{rx},\psiv}^* \RB (t) &= 
	\begin{cases}
	\frac{\sqrt{C(\thetav)}}{D} \psiv\htp\Am(t)\thetav & \text{if} \abs{t} \leq D\\
	0 & \text{otherwise}
	\end{cases}	
	\label{eq:tx_rx_conv}\\
	\EE \LSB w_m w_{m+\ell}^* \RSB &=
	\begin{cases}
	\frac{N_0}{D} \psiv\htp \Am'\LB \ell T \RB \psiv & \text{if} \abs{\ell T} \leq D\\
	0 & \text{otherwise}
	\end{cases} \label{eq:noise_conv_filt}
\end{align}
where $\Am(t)\in\CC^{(2S+1)\times(2S+1)}$ has elements
\begin{multline}
	A(t)_{s_1,s_2} =\\
	\begin{cases}
	e^{j2\pi\frac{s_2}{D}t}\Delta(t) &\text{if } s_1 + s_2 = 0\\
	e^{j\pi\LB 2\frac{s_2}{D}t - (s_1+s_2)\Sc(t)\RB}\frac{\sin \LB \pi(s_1+s_2)\Delta(t) \RB}{\pi(s_1+s_2)} &\text{otherwise}
	\end{cases}
\end{multline}
where $-S \leq s_1,s_2 \leq S$, $\Delta(t) = L_{max}(t) - L_{min}(t)$, and $\Sc(t) = L_{max}(t) + L_{min}(t)$, with $L_{max}(t) = \min\LP \frac{1}{2} ; \frac{t}{D} + \frac{1}{2} \RP$, and $L_{min}(t) = \max\LP -\frac{1}{2} ; \frac{t}{D} - \frac{1}{2} \RP$.
Similarly, $\Am'(t)\in\CC^{(2S+1)\times(2S+1)}$ has elements
\begin{multline}
	A'(t)_{s_1,s_2} =\\
	\begin{cases}
	e^{j2\pi\frac{s_1}{D}t}\Delta'(t) &\text{ if } s_1 = s_2\\
	e^{j\pi\LB 2\frac{s_1}{D}t + (s_2-s_1)\Sc'(t)\RB}\frac{\sin \LB \pi(s_2-s_1)\Delta'(t) \RB}{\pi(s_2 - s_1)} &\text{ otherwise}
	\end{cases}
\end{multline}
where $-S \leq s_1,s_2 \leq S$, $\Delta'(t) = L_{max}'(t) - L_{min}'(t)$, and $\Sc'(t) = L_{max}'(t) + L_{min}'(t)$, with $L_{max}'(t) = \min\LP \frac{1}{2} ; -\frac{t}{D} + \frac{1}{2} \RP$, and $L_{min}'(t) = \max\LP -\frac{1}{2} ; -\frac{t}{D} - \frac{1}{2} \RP$.
Detailed derivations of~\eqref{eq:c_norm}, \eqref{eq:tx_rx_conv}, and \eqref{eq:noise_conv_filt} are provided in the Appendix.

\subsection{Trainable constellation}
\label{sec:const_tr}

As done, e.g., in~\cite{aitaoudia2020}, the trainable constellation consists of a set of $2^K$ complex numbers denoted by $\tilde{\Cc}$, which represent the uncentered and unnormalized constellation points.
To ensure that the constraint~\eqref{eq:pc2} is satisfied, $\tilde{\Cc}$ is centered and normalized as
\begin{equation}
	\label{eq:const_norm}
	\Cc = \frac{\tilde{\Cc} - 2^{-K}\sum_{c\in\tilde{\Cc}}c}{\sqrt{2^{-K}\sum_{c\in\tilde{\Cc}}\abs{c}^2 - \abs{2^{-K}\sum_{c\in\tilde{\Cc}}c}^2}}.
\end{equation}
Although centering is not strictly necessary,  it avoids a possibly unwanted \gls{DC} offset.
The centered and normalized constellation $\Cc$ is then used to modulate the data bits.
Each point of $\tilde{\Cc}$ (and $\Cc$) has a predefined arbitrary bit label.
By training on the \gls{BMD} rate~\eqref{eq:R}, optimization of the constellation geometry is performed while considering the labeling of each point. This leads to the joint optimization of the bit labeling and constellation geometry.
Note that no additional complexity is introduced on the transmitter side compared to a traditional single-carrier system since it relies on \gls{BICM} using the constellation $\Cc$~\eqref{eq:const_norm} and pulse shaping using the filter in \eqref{eq:tx}.
Once the end-to-end system is trained, the learned transmit filter, constellation geometry, and corresponding labeling can simply be used \emph{in lieu} of, e.g., an \gls{RRC} filter and \gls{QAM} constellation with Gray labeling.

\subsection{Neural network receiver}
\label{sec:nn_tr}

\begin{figure*}
	\begin{subfigure}[b]{0.3\linewidth}
		\centering
		\footnotesize
		\def\svgwidth{1\linewidth}
\begingroup%
  \makeatletter%
  \providecommand\color[2][]{%
    \errmessage{(Inkscape) Color is used for the text in Inkscape, but the package 'color.sty' is not loaded}%
    \renewcommand\color[2][]{}%
  }%
  \providecommand\transparent[1]{%
    \errmessage{(Inkscape) Transparency is used (non-zero) for the text in Inkscape, but the package 'transparent.sty' is not loaded}%
    \renewcommand\transparent[1]{}%
  }%
  \providecommand\rotatebox[2]{#2}%
  \newcommand*\fsize{\dimexpr\f@size pt\relax}%
  \newcommand*\lineheight[1]{\fontsize{\fsize}{#1\fsize}\selectfont}%
  \ifx\svgwidth\undefined%
    \setlength{\unitlength}{230.54617093bp}%
    \ifx\svgscale\undefined%
      \relax%
    \else%
      \setlength{\unitlength}{\unitlength * \real{\svgscale}}%
    \fi%
  \else%
    \setlength{\unitlength}{\svgwidth}%
  \fi%
  \global\let\svgwidth\undefined%
  \global\let\svgscale\undefined%
  \makeatother%
  \begin{picture}(1,0.42940938)%
    \lineheight{1}%
    \setlength\tabcolsep{0pt}%
    \put(0,0){\includegraphics[width=\unitlength]{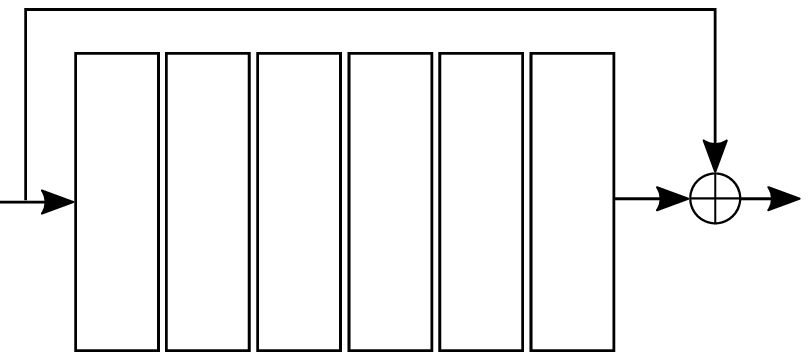}}%
    \put(0.15789895,0.05654999){\color[rgb]{0,0,0}\rotatebox{90}{\makebox(0,0)[lt]{\lineheight{1.25}\smash{\begin{tabular}[t]{l}Batch norm.\end{tabular}}}}}%
    \put(0.27073892,0.11503292){\color[rgb]{0,0,0}\rotatebox{90}{\makebox(0,0)[lt]{\lineheight{1.25}\smash{\begin{tabular}[t]{l}ReLU\end{tabular}}}}}%
    \put(0.38208659,0.02156638){\color[rgb]{0,0,0}\rotatebox{90}{\makebox(0,0)[lt]{\lineheight{1.25}\smash{\begin{tabular}[t]{l}Separable Conv.\end{tabular}}}}}%
    \put(0.49923452,0.05027657){\color[rgb]{0,0,0}\rotatebox{90}{\makebox(0,0)[lt]{\lineheight{1.25}\smash{\begin{tabular}[t]{l}Batch norm.\end{tabular}}}}}%
    \put(0.61207449,0.12177209){\color[rgb]{0,0,0}\rotatebox{90}{\makebox(0,0)[lt]{\lineheight{1.25}\smash{\begin{tabular}[t]{l}ReLU\end{tabular}}}}}%
    \put(0.7234222,0.02179925){\color[rgb]{0,0,0}\rotatebox{90}{\makebox(0,0)[lt]{\lineheight{1.25}\smash{\begin{tabular}[t]{l}Separable Conv.\end{tabular}}}}}%
  \end{picture}%
\endgroup%

    	\caption{Architecture of a ResNet block.}
    	\label{fig:resnet_block}
   	\end{subfigure}
	\quad
    \begin{subfigure}[b]{0.7\linewidth}
    	\centering
		\footnotesize
    	\def\svgwidth{0.9\linewidth}
		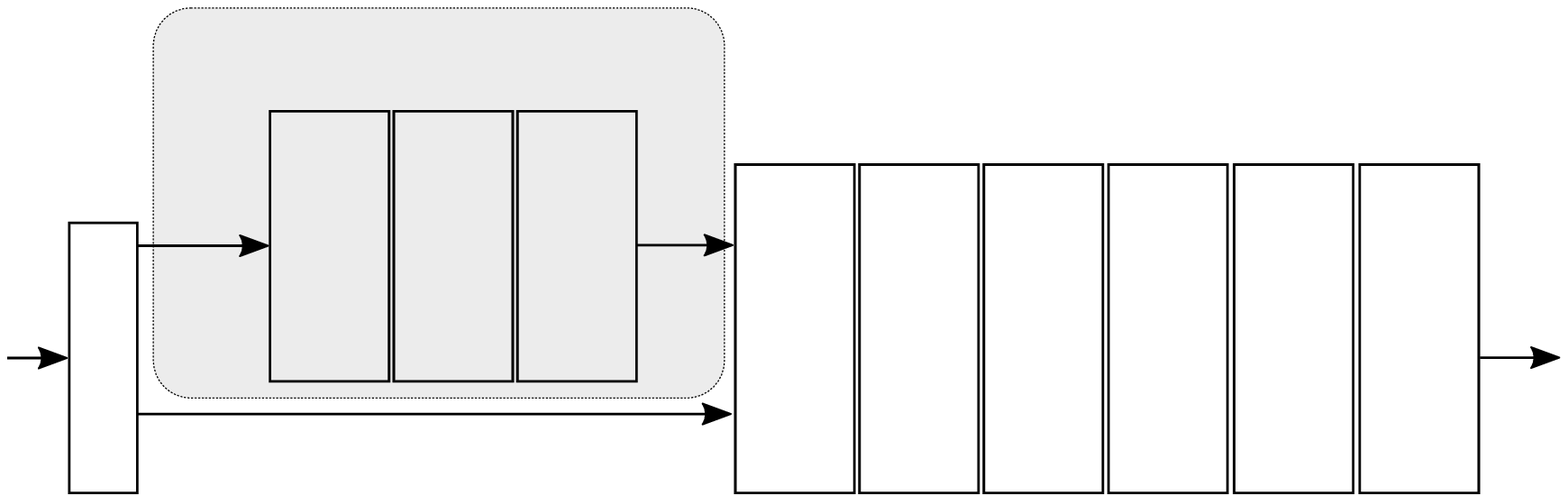    	
    	\caption{\gls{NN}-based detector for the \gls{AWGN} and multipath channels.}
    	\label{fig:nnrx}
    \end{subfigure}

     \caption{Architecture of the neural receivers for the \gls{AWGN} and multipath channels. For each ResNet block (dense layer), the number of kernels, kernel size, and dilatation rate (activation and number of units) are indicated.}
\end{figure*}

The neural receiver computes \glspl{LLR} on the transmitted bits $\Bm$ from the received samples $\rv \in \CC^N$.
As shown in Fig.~\ref{fig:nnrx}, the neural receiver for the \gls{AWGN} is a residual 1-dimensional convolutional \gls{NN}~\cite{He_2016_CVPR}. For multipath channels, an additional dense \gls{NN} is needed to extract channel state information. This will be explained in more detail in Section~\ref{sec:mp}.

The convolutional \glspl{NN} are made of several ResNet blocks that have the generic architecture shown in Fig.~\ref{fig:resnet_block}.
Separable convolutional layers are used to reduce the complexity and zero-padding is employed to ensure that the output has the same length as the input.
We use dilatation to increase the receptive field of the convolutional layers.
As shown in Fig.~\ref{fig:nnrx}, the first layer is a $\CC2\RR$ layer which converts the complex-valued input of length $N$ to a 2-dimensional real-valued tensor with shape $N \times 2$, by stacking the real and imaginary parts.
From a deep learning perspective, the second dimension serves as the ``channels'' of the convolutional \gls{NN}.
The output of the neural receiver is a two-dimensional real-valued tensor with shape $N \times K$, which corresponds to the \glspl{LLR} of the transmitted bits $\Bm$.

For the multipath channel, it is assumed that the first $N_P$ symbols of each block are pilots. As shown in Fig.~\ref{fig:nnrx},
these are processed by a dense \gls{NN} which outputs a real-valued vector $\zv$ of size $N_S$.
The vector $\zv$ is then stacked together with the real and imaginary parts of the received symbols $\rv$ along the ``channels'' dimension, to form a real-valued tensor with dimension $N \times (N_S + 2)$ that serves as input to the convolutional neural receiver. 
This architecture was designed with the intuition that $\zv$ carries information on the channel state which is helpful for computing the \glspl{LLR}. We were not able to train the system without $\zv$ and dedicated pilots.

\subsection{Training process}
\label{sec:tr_proc}

Training of the end-to-end system consists in jointly optimizing the transmit and receive filters, the neural receiver, the constellation as well as the corresponding bit labeling to solve~\eqref{eq:prob_main}.
The maximization of the achievable rate~\eqref{eq:R} is achieved by minimizing the total binary-cross-entropy
\begin{multline}
	\label{eq:loss}
	\Lc(\Cc,\thetav,\psiv,\gammav)\\
	\coloneqq -\frac{1}{N} \sum_{n=0}^{N-1} \sum_{k=0}^{K-1} \EE \LSB \log{Q_{\gammav}\LB B_{n,k} | \rv \RB} | \Cc,\thetav,\psiv \RSB
\end{multline}
which is related to $R$ by
\begin{equation}
	\Lc(\Cc, \thetav,\psiv,\gammav) = K - R(\Cc,\thetav,\psiv,\gammav).
\end{equation}
Since~\eqref{eq:loss} is numerically difficult to compute, it is estimated through Monte Carlo sampling:
\begin{multline}
	\Lc(\Cc,\thetav,\psiv,\gammav)\\
	\approx -\frac{1}{MN} \sum_{m=0}^{M-1} \sum_{n=0}^{N-1} \sum_{k=0}^{K-1} \log{Q_{\gammav}\LB B_{n,k}^{[m]} \lvert \rv^{[m]} \RB}
\end{multline}
where $M$ is the batch size, i.e., the number of samples used to compute the estimate of $\Lc$, and the superscript $[m]$ is used to refer to the $m^{th}$ example within a batch.

Finding a local optimum of~\eqref{eq:prob} is made challenging by the constraints~\eqref{eq:pc1}--\eqref{eq:pc4}.
The constraint~\eqref{eq:pc1} is guaranteed through the normalization constant $C(\thetav)$~\eqref{eq:c_norm} in~\eqref{eq:tx}, and the constraint~\eqref{eq:pc2} is achieved through the normalization of the constellation~\eqref{eq:const_norm}.
Enforcing the constraint on the \gls{ACLR}~\eqref{eq:pc3}, however, requires the computation of the in-band energy~\eqref{eq:inb} as it determines the \gls{ACLR}~\eqref{eq:aclr}.
With the implementation of the trainable transmit filter introduced in Section~\ref{sec:tr_filters}, the in-band energy can be efficiently computed as
\begin{equation}
	\label{eq:ei_trf}
	E_I(\thetav) = C(\thetav) \thetav\htp \Em \thetav
\end{equation}
where $\Em\in\RR^{(2S+1)\times(2S+1)}$ has elements
\begin{equation}
	E_{s_1,s_2} = \int_{-\frac{W}{2}}^{\frac{W}{2}} \text{sinc}\LB Df - s_1 \RB \text{sinc}\LB Df - s_2 \RB df
\end{equation}
which can be pre-computed prior to training.
This leads to
\begin{equation}
	\text{ACLR}(\thetav) = \frac{1}{C(\thetav) \thetav\htp \Em \thetav}-1
\end{equation}
when combined with~\eqref{eq:aclr}.
A derivation of~\eqref{eq:ei_trf} is provided in the Appendix.

Concerning the constraint on the \gls{PAPR}~\eqref{eq:pc4}, we use the equivalent constraint function~\eqref{eq:pc4_equiv} and estimate the quantity $V$ by Monte Carlo sampling
\begin{equation}
	\label{eq:v_approx}
	V(\thetav, \Cc, \epsilon_P) \approx \frac{1}{M'} \sum_{m=1}^{M'} \max \LB \frac{p^{[m]}(t^{[m]})}{\bar{p}} - \epsilon_P, 0 \RB
\end{equation}
where $M'$ is the number of samples. The signal power samples $p^{[m]}(t^{[m]})$ are generated by sampling first random baseband symbols $\sv$ from the constellation $\Cc$, which are used to compute the time-continuous signal $x(t)$ according to \eqref{eq:x}. The power $p(t) = \abs{x(t)}^2$ is then sampled at randomly chosen instants $t^{[m]}\sim \Uc\LB -\frac{T}{2}, \frac{T}{2} \RB$.

\begin{algorithm}
\caption{Training algorithm}
\label{alg:alt}
\begin{algorithmic}[1]
\State Initialize $\tilde{\Cc}$, $\thetav$, $\psiv$, $\gammav$, $\eta^{[0]}$, $\lambda_A^{[0]}$, and $\lambda_P^{[0]}$.
  \For{$u=0,\cdots$}
    \State Perform SGD on $\Lc_A\LB \Cc,\thetav,\psiv,\gammav;\lambda_A^{[u]},\lambda_P^{[u]}, \eta^{[u]} \RB$\label{lst:sgd}
    \State \LineComment{Update Lagrange multipliers:}
    \State $\lambda_A^{[u+1]} \gets \lambda_A^{[u]} - \eta^{[u]} \max \LB \text{ACLR}(\thetav) - \epsilon_A, 0 \RB$
    \State $\lambda_P^{[u+1]} \gets \lambda_P^{[u]} - \eta^{[u]} V(\thetav,\Cc,\epsilon_P)$
    \State \LineComment{Update penalty parameter:}
    \State Set $\eta^{[u+1]}$ such that $\eta^{[u+1]} > \eta^{[u]}$ \label{lst:eta}
  \EndFor
\end{algorithmic}
\end{algorithm}

Finding a local maximum of~\eqref{eq:prob_main} is achieved by using the augmented Lagrangian method~\cite[Chapter~3]{bertsekas2014constrained}, which is shown in Algorithm~\ref{alg:alt} when applied to our setup.
The augmented Lagrangian method consists in solving a sequence of unconstrained optimization problems (line~\ref{lst:sgd}), each aiming at minimizing the augmented Lagrangian
\begin{multline}
	\Lc_A \LB \Cc,\thetav,\psiv,\gammav;\lambda_A^{[u]},\lambda_P^{[u]}, \eta^{[u]} \RB \coloneqq \Lc(\Cc,\thetav,\psiv,\gammav)\\
	- \lambda_P^{[u]} V(\thetav,\Cc, \epsilon_P) - \lambda_A^{[u]} \max \LB \text{ACLR}(\thetav) - \epsilon_A, 0 \RB\\
	+ \frac{\eta}{2}\LB V(\thetav,\Cc, \epsilon_P)^2 + \LB \max \LB \text{ACLR}(\thetav) - \epsilon_A, 0 \RB \RB^2 \RB
\end{multline} 
where the superscript $[u]$ refers to the $u^{th}$ iteration, $\lambda_A$ and $\lambda_P$ are the Lagrange multipliers, and $\eta$ is a positive penalty parameter that is progressively increased (line~\ref{lst:eta}).
At each iteration, minimizing the augmented Lagrangian is approximately achieved through \gls{SGD}.

\section{Simulation results}
\label{sec:sr}

\subsection{Setup}

The multipath channel model implemented as described in Section~\ref{sec:sm} was considered to train and benchmark the end-to-end learning approach.
For brevity, the baseline is referred to as BL, and the end-to-end learning approach as E2E.
The carrier frequency was set to $3.5\:$GHz and the bandwidth to $W = 5\:$MHz.
The duration of the transmit and receive filters was set to $D = 32T$ and their number of trainable parameters was set to $S = 100$.
The block length was set to $N = 1000$ symbols and the modulation order to $16$ ($K = 4$).
The \gls{NN}-based receiver operates on the entire block of $N$ symbols.
The \gls{SNR} is defined as
\begin{equation}
\text{SNR} \coloneqq \frac{1}{N_0}.
\end{equation}

Training of the E2E system was carried out using the Adam optimizer~\cite{adam} to perform \gls{SGD}, with batches of size $M = 10$ and a learning rate of $10^{-3}$.
The batch size for evaluating the \gls{PAPR} constraint~\eqref{eq:v_approx} was set to $M' = 10000$.
The penalty parameter was initialized with $\eta^{[0]} = 10^{-2}$ and then increased according to  a multiplicative schedule $\eta^{[u+1]} = 1.003\eta^{[u]}$.
The Lagrange multipliers were both initialized with $0$.

\subsection{\gls{AWGN} channel}
\label{sec:awgn}

\begin{figure}[h]
\centering{
\begin{tikzpicture}

\definecolor{color0}{rgb}{0.12156862745098,0.466666666666667,0.705882352941177}
\definecolor{color1}{rgb}{1,0.498039215686275,0.0549019607843137}
\definecolor{color2}{rgb}{0.172549019607843,0.627450980392157,0.172549019607843}
\definecolor{color3}{rgb}{0.83921568627451,0.152941176470588,0.156862745098039}
\definecolor{color4}{rgb}{0.580392156862745,0.403921568627451,0.741176470588235}

\begin{axis}[
legend cell align={left},
legend style={
  fill opacity=0.8,
  draw opacity=1,
  text opacity=1,
  at={(0.7,0.01)},
  anchor=south west,
  draw=white!80!black
},
tick align=outside,
tick pos=left,
x grid style={white!69.0196078431373!black},
xlabel={ACLR [\si{\decibel}]},
xmajorgrids,
xmin=-55, xmax=0,
xtick style={color=black},
y grid style={white!69.0196078431373!black},
ylabel={\(\displaystyle R\) [\si{\bit\per\second\per\hertz}]},
ymajorgrids,
ymin=2.05, ymax=3.45,
ytick style={color=black},
ytick={2,2.2,2.4,2.6,2.8,3,3.2,3.4,3.6},
yticklabels={2.0,2.2,2.4,2.6,2.8,3.0,3.2,3.4,3.6}
]
\addplot [semithick, color0, mark=*, mark size=3, mark options={solid}]
table {%
-21.4701771736145 3.0494978427887
-13.2707405090332 3.16188955307007
-10.0144100189209 3.16323328018188
-8.02195727825165 3.16334795951843
-6.53488218784332 3.16337132453918
};
\addlegendentry{BL}
\addplot [semithick, color1, mark=square*, mark size=3, mark options={solid}]
table {%
-52.3349380493164 2.14117503166199
-41.6477823257446 2.54821920394897
-30.0807905197144 2.58648467063904
-19.785168170929 2.78577518463135
};
\addlegendentry{E2E $\epsilon_P = \SI{4}{\decibel}$}
\addplot [semithick, color2, mark=diamond*, mark size=3, mark options={solid}]
table {%
-49.8417043685913 2.71570611000061
-40.3551054000854 2.77044439315796
-29.7186684608459 2.85942316055298
-19.8188543319702 3.056964635849
};
\addlegendentry{E2E $\epsilon_P = \SI{5}{\decibel}$}
\addplot [semithick, color3, mark=triangle*, mark size=3, mark options={solid}]
table {%
-50.1520490646362 2.93625736236572
-40.5095100402832 2.98768758773804
-30.0039410591125 3.05108666419983
-20.1059699058533 3.15338659286499
};
\addlegendentry{E2E $\epsilon_P = \SI{6}{\decibel}$}
\addplot [semithick, color4, mark=triangle*, mark size=3, mark options={solid,rotate=180}]
table {%
-49.6464872360229 2.99764132499695
-40.0408887863159 3.06036114692688
-30.1767826080322 3.11452627182007
-20.1753282546997 3.15614008903503
};
\addlegendentry{E2E $\epsilon_P = \SI{7}{\decibel}$}
\draw (axis cs:-22,3.1) node[
  scale=0.6,
  anchor=north east,
  text=color0,
  rotate=0.0
]{$0.0$};
\draw (axis cs:-3,3.28) node[
  scale=0.6,
  anchor=north east,
  text=color0,
  rotate=0.0
]{$1.0$};
\draw (axis cs:-43,2.17) node[
  scale=0.6,
  anchor=north east,
  text=color1,
  rotate=0.0
]{\SI{-50}{\decibel}};
\draw (axis cs:-36,2.5) node[
  scale=0.6,
  anchor=north east,
  text=color1,
  rotate=0.0
]{\SI{-40}{\decibel}};
\draw (axis cs:-25,2.56) node[
  scale=0.6,
  anchor=north east,
  text=color1,
  rotate=0.0
]{\SI{-30}{\decibel}};
\draw (axis cs:-14,2.74) node[
  scale=0.6,
  anchor=north east,
  text=color1,
  rotate=0.0
]{\SI{-20}{\decibel}};
\end{axis}

\end{tikzpicture}
\caption{Rates and \gls{ACLR} achieved by the BL and E2E systems. The different points for the BL (from left to right) are obtained by having the roll-off factor $\beta$ take values from the set $\{0.0, 0.25, 0.50, 0.75, 1.0\}$. For the E2E system, values of $\epsilon_A$ from the set $\{-50, -40, -30, -20\}\:$dB (from left to \textcolor{black}{right}) were considered. These are only indicated for $\epsilon_P = 4.0\:$dB for clarity.}
\label{fig:rate_aclr}
}
\end{figure}
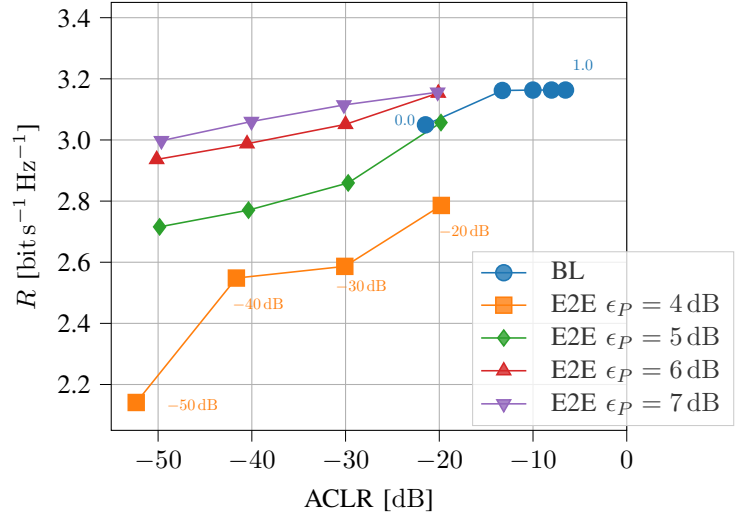

An \gls{AWGN} channel with an \gls{SNR} of \SI{10}{\decibel} is considered in this section.
More precisely, in the transfer function~\eqref{eq:h}, the number of paths is set to $P = 1$, with $a_0 = 1$ and $\tau_0 = 0$.
The E2E system uses the neural receiver shown in Fig.~\ref{fig:nnrx}.
A Gray-labeled 16\gls{QAM} was used as BL with the windowed \gls{RRC} filter as introduced in Section~\ref{sec:bsl}.

Fig.~\ref{fig:rate_aclr} shows the rates $R$ and \glspl{ACLR} achieved by the BL and the E2E schemes, for different values of $\epsilon_A$ and $\epsilon_P$. 
Concerning the BL, values of the roll-off factor $\beta$ from the set $\{0.0, 0.25, 0.50, 0.75, 1.0\}$ were considered.
As shown in Fig~\ref{fig:rate_aclr}, setting the roll-off factor to 0 gives the lowest \gls{ACLR}, at the cost of a rate loss due to the \gls{ISI} introduced by the windowing.
Higher values of $\beta$ lead to lower ripples which make the filter more robust to windowing, at the cost of higher \gls{ACLR}.
Concerning the E2E approach, values of $\epsilon_A$ from the set $\{-50, -40, -30, -20\}\:$\si{\decibel} were considered, and are indicated in Fig.~\ref{fig:rate_aclr} only for the plot corresponding to $\epsilon_P = \SI{4}{\decibel}$ for clarity.
As one can see, the \gls{ACLR} incurred by the E2E system is always lower or very close to the value of $\epsilon_A$ for which it was trained. This indicates that the training process allows accurate control of the \gls{ACLR} through this parameter.
Moreover, it enables \glspl{ACLR} up to \SI{30}{\decibel} lower than the ones of the BL, without significant rate loss.

\begin{figure}[h]
\centering{
\input{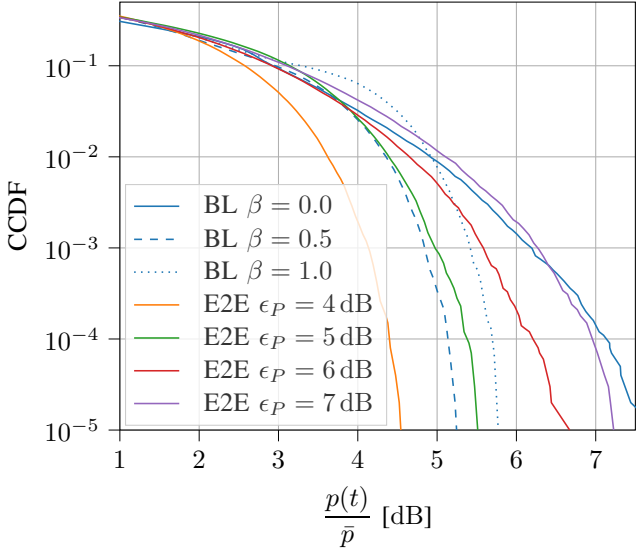}
\caption{CCDF of the normalized power of the BL and E2E system for $\epsilon_A = \SI{-30}{\decibel}$.}
\label{fig:ccdf}
}
\end{figure}

\begin{figure}[h]
\centering{
\input{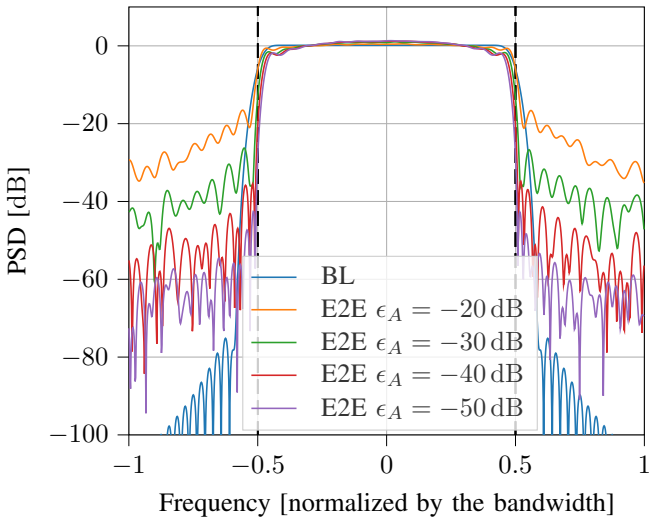}
\caption{\glspl{PSD} of the baseline ($\beta = 0$) and the E2E system for $\epsilon_P = \SI{6}{\decibel}$.}
\label{fig:psd}
}
\end{figure}

The \gls{PAPR} constraint has the strongest impact on the rate.
To quantify the impact of the \gls{PAPR} constraint on the power distribution, Fig.~\ref{fig:ccdf} shows the \gls{CCDF} of the normalized power for $\epsilon_A = \text{\SI{10}{\decibel}}$.
As one can see, constraining the \gls{PAPR} has the expected impact on the power distribution, as it significantly reduces the occurrences of power peaks, at the cost of a rate loss (Fig.~\ref{fig:rate_aclr}).
The E2E system achieves the lowest power peaks when $\epsilon_P = \SI{4}{\decibel}$, whereas setting $\epsilon_P$ to \SI{7}{\decibel} leads to the same \gls{CCDF} as for a windowed sinc filter ($\beta = 0$).
Concerning the power spectrum, Fig.~\ref{fig:psd} shows the \glspl{PSD} of the learned filters ($\epsilon_P = \SI{6}{\decibel}$) and that of the baseline ($\beta = 0$).
One can see that the baseline emits more in the adjacent bands, i.e., near the dashed vertical black lines that define the in-band region, whereas the learned waveforms leak less in the adjacent bands, but reach higher leakage floors.
Nevertheless, the \gls{ACLR} values are better because these floors contribute only little to the total out-of-band emissions.
In practice, the minimization of the leakage adjacent to the in-band matters more since the emission floor is typically below the noise level.

\begin{figure*}
\centering{
	\scalebox{.6}{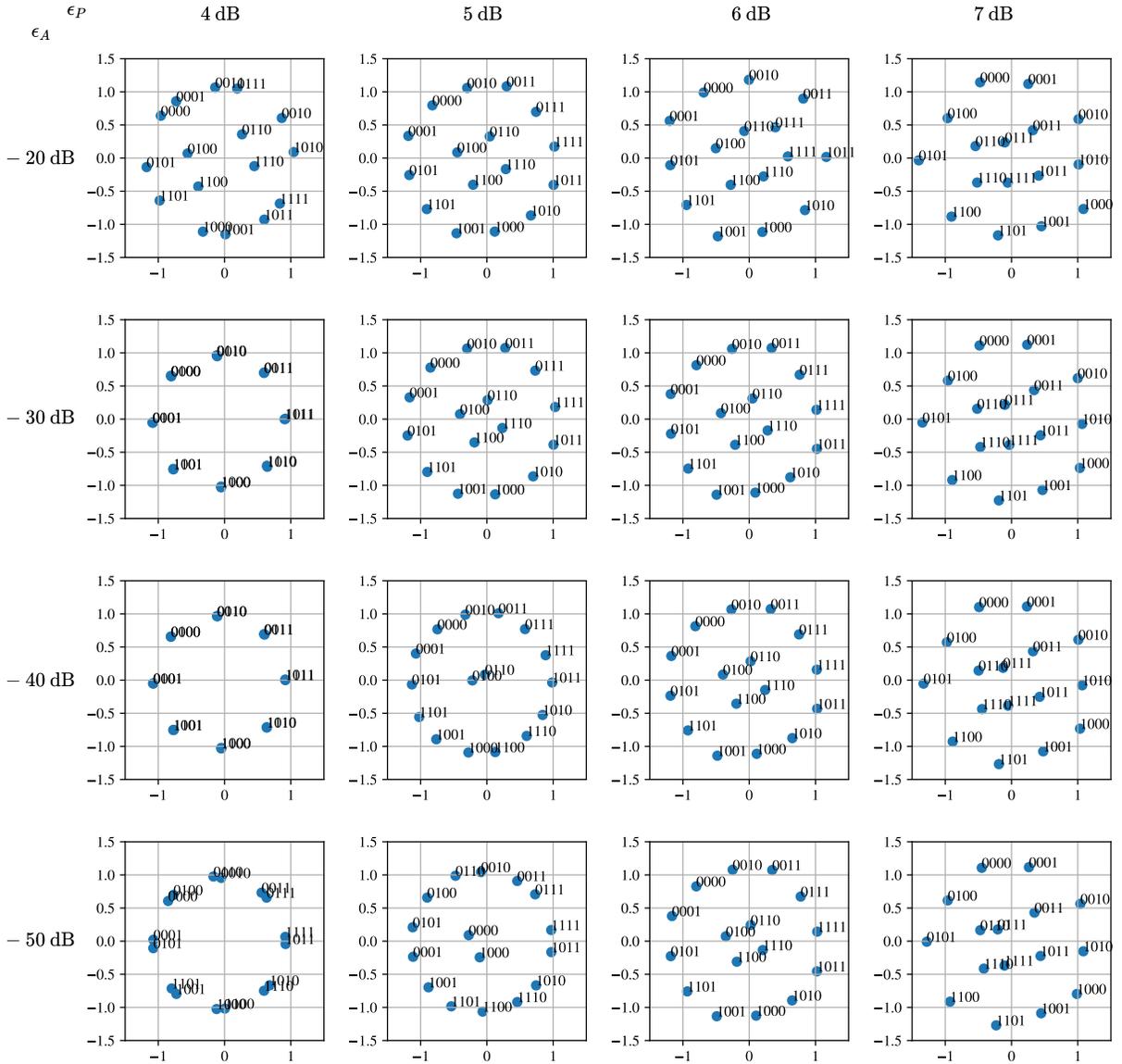}
	\caption{Learned constellation geometries and bit labels for different values of $\epsilon_A$ and $\epsilon_P$.}
	\label{fig:awgn_consts}
}
\end{figure*}

\begin{figure*}
\centering{
	\scalebox{.5}{
\begingroup%
  \makeatletter%
  \providecommand\color[2][]{%
    \errmessage{(Inkscape) Color is used for the text in Inkscape, but the package 'color.sty' is not loaded}%
    \renewcommand\color[2][]{}%
  }%
  \providecommand\transparent[1]{%
    \errmessage{(Inkscape) Transparency is used (non-zero) for the text in Inkscape, but the package 'transparent.sty' is not loaded}%
    \renewcommand\transparent[1]{}%
  }%
  \providecommand\rotatebox[2]{#2}%
  \newcommand*\fsize{\dimexpr\f@size pt\relax}%
  \newcommand*\lineheight[1]{\fontsize{\fsize}{#1\fsize}\selectfont}%
  \ifx\svgwidth\undefined%
    \setlength{\unitlength}{894.65631104bp}%
    \ifx\svgscale\undefined%
      \relax%
    \else%
      \setlength{\unitlength}{\unitlength * \real{\svgscale}}%
    \fi%
  \else%
    \setlength{\unitlength}{\svgwidth}%
  \fi%
  \global\let\svgwidth\undefined%
  \global\let\svgscale\undefined%
  \makeatother%
  \begin{picture}(1,0.85214397)%
    \lineheight{1}%
    \setlength\tabcolsep{0pt}%
    \put(0,0){\includegraphics[width=\unitlength]{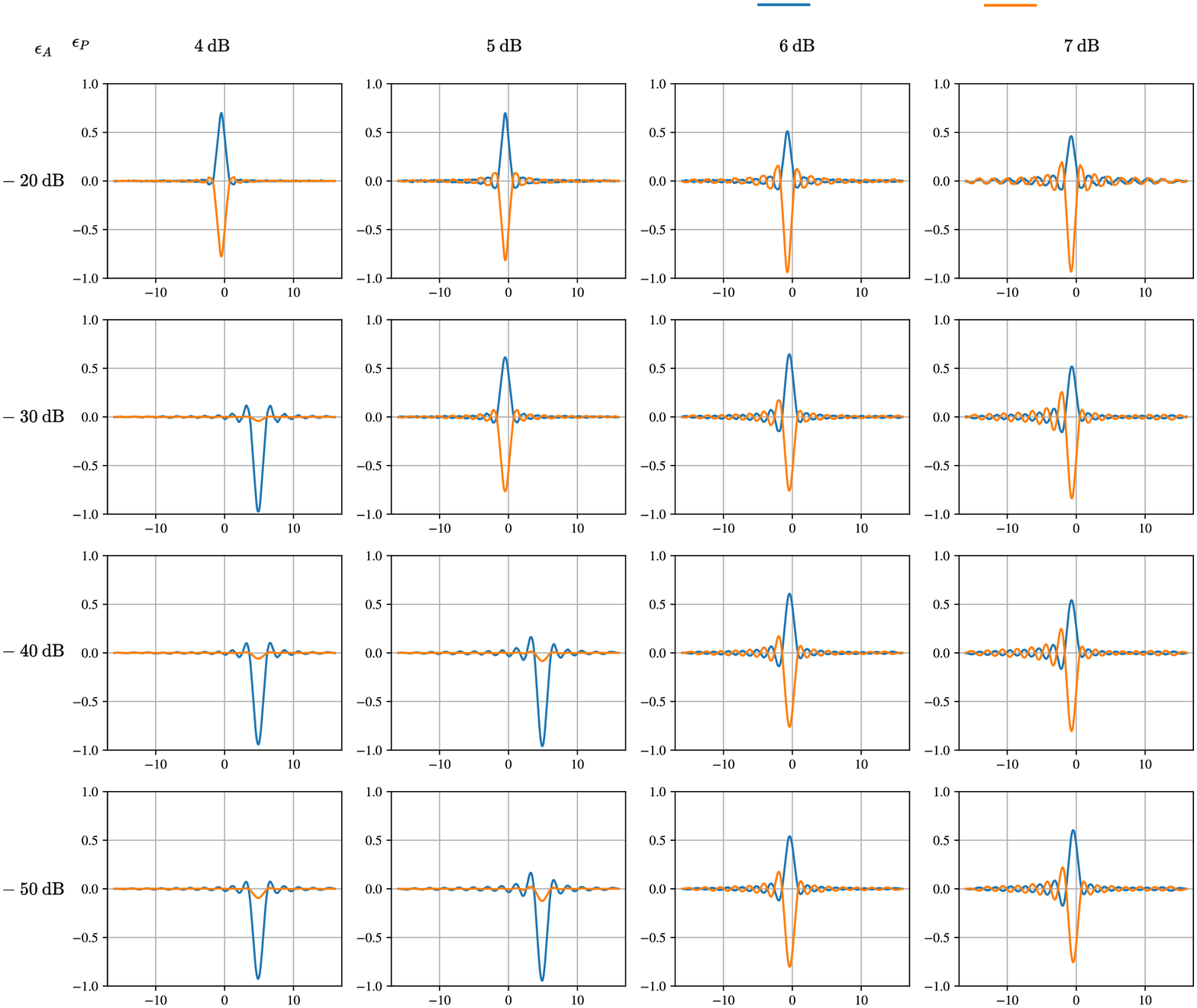}}%
    \put(0.68536339,0.83685789){\color[rgb]{0,0,0}\makebox(0,0)[lt]{\lineheight{1.25}\smash{\begin{tabular}[t]{l}Real part\end{tabular}}}}%
    \put(0.8756726,0.83636563){\color[rgb]{0,0,0}\makebox(0,0)[lt]{\lineheight{1.25}\smash{\begin{tabular}[t]{l}Imag. part\end{tabular}}}}%
  \end{picture}%
\endgroup%
}
	\caption{Learned transmit filters for different values of $\epsilon_A$ and $\epsilon_P$.}
	\label{fig:awgn_tps}
}
\end{figure*}

To get some more insight into the E2E system, we show in Fig.~\ref{fig:awgn_consts} and Fig.~\ref{fig:awgn_tps} the learned constellations and transmit filters, respectively, for different values of $\epsilon_A$ and $\epsilon_P$. As one can see from Fig.~\ref{fig:awgn_consts}, a form of Gray labeling is always learned jointly with the constellation geometry.
When strongly constraining both the \gls{ACLR} and \gls{PAPR} (bottom left), a \gls{PSK} constellation is learned and points are clustered by groups of two which differ only in a single bit. This means that the E2E system is willingly ``sacrificing'' a bit of information to improve the \gls{PAPR}.
Note that the modulation order could therefore be reduced to eight for these setups.
Lowering either of the constraints leads to constellations with two levels of amplitudes.
The softest constraints (top row) lead to the least intuitive geometries. Overall, as expected, it seems that the constellations are mainly determined by the \gls{PAPR} and not the \gls{ACLR} constraint.

Interpreting the learned transmit filters in Fig.~\ref{fig:awgn_tps} is less straightforward.
The learned filters have some similarity with an \gls{RRC} filter, especially when both constraints are strong (bottom left). In this case, the imaginary component vanishes, similar to conventional filters that have only a real component.
However, relaxing the constraints leads to significant imaginary components.
Moreover, the pulse shape is not symmetric around the peak in contrast to conventional filters. In contrast to the constellation geometry, which is mainly determined by the \gls{PAPR} constraint, the pulse shape is---a bit unexpectedly---impacted by both constraints. 

\subsection{Multiple access channel}
\label{sec:awgn_mac}

\begin{figure}[h]
\small
\centering{
\def\svgwidth{\linewidth}
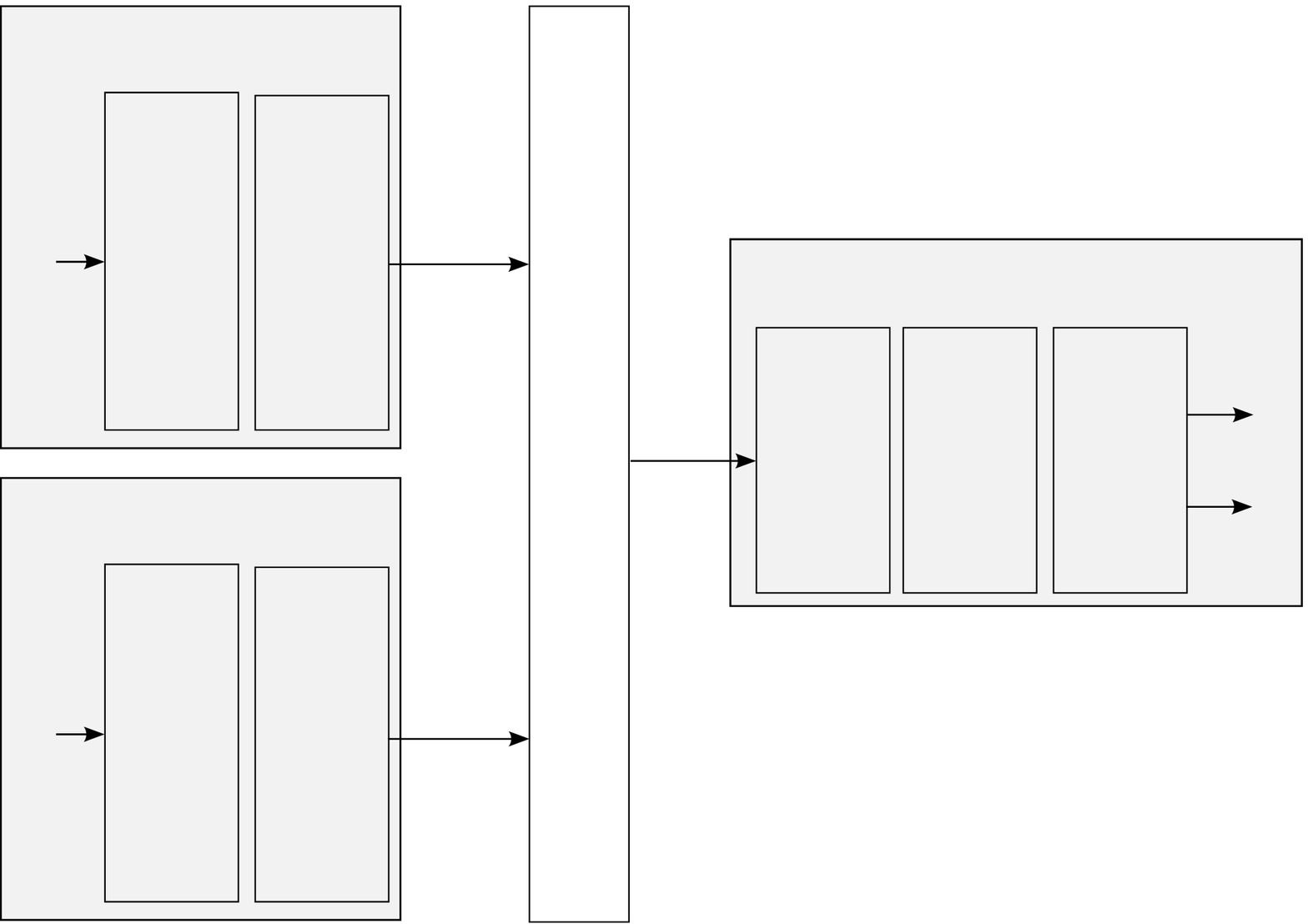
\caption{End-to-end system with two interfering users, each with a trainable transmit filter and constellation, and a receiver with a trainable receive filter and NN.}
\label{fig:sys_mac}
}
\end{figure}

User multiplexing is typically achieved through \gls{OMA} in the time, frequency, code, or spatial domains.
For example, different users would share the available spectrum by transmitting only on dedicated bands, using the \gls{ACLR} as a measure of the power a user leaks onto the adjacent bands used by the other users.
It is well-known that such \gls{OMA} approaches are sub-optimal and a significant research effort is dedicated to the design of \gls{NOMA} schemes~\cite{8972353}.
In this section, we explore how the use of dedicated constellation geometries and transmit filters could enable multiplexing.
An \gls{AWGN} channel with \gls{SNR} set to \SI{10}{\decibel} is considered, as in Section~\ref{sec:awgn}.
However, we now consider two interfering users as illustrated in Fig.~\ref{fig:sys_mac}.
The two users transmit with the same average power and over the same band.
Each user has a dedicated trainable constellation geometry, bit labeling, and transmit filter, as shown in Fig.~\ref{fig:sys_mac}.
The receiver includes a trainable receive filter and neural receiver, that are jointly optimized together with the two trainable transmitters over the interfering channel.

Training is done on the sum-log rate
\begin{equation}
	\Rc \coloneqq w\log{R_1} + (1-w)\log{R_2}
\end{equation}
where $R_1$ and $R_2$ are the achievable \gls{BMD} rates~\eqref{eq:R} for user~1 and user~2, respectively, and
$w \in (0,1)$ allows favoring one user over the other.
Intuitively, by using the sum-log rate, increasing the rate of the user with the highest rate leads to less reward than increasing the rate of the user with the lowest rate, thus favoring fairness.
Formally, the sum-log rate is closely related to proportionality fairness~\cite{ett.4460080106}.
A rate allocation $(r_1, r_2)$ is said to be proportionally fair if for any other allocation $(r_1', r_2')$, one has
\begin{equation}
w\frac{r_1' - r_1}{r_1} + (1-w)\frac{r_2' - r_2}{r_2} \leq 0.
\end{equation}
Especially, when $w = 0.5$, increasing the rate of one user by $p\:\si{\percent}$ can only be achieved by reducing the rate of the other user by at least $p\:\si{\percent}$.
Except for the different loss function, training of the end-to-end system was achieved using the method depicted in Section~\ref{sec:tr_proc}, with $\epsilon_A$ set to \SI{-30}{\decibel} and $\epsilon_P$ set to \SI{6}{\decibel}.
The two users are subject to the same \gls{PAPR} and \gls{ACLR} constraints.
The latter enforces them explicitly to operate on the same band.

\begin{figure}[h]
\centering
\input{figs/rates_mac.tex}
\caption{Achievable rates for the two interfering users, with $\epsilon_A = \SI{-30}{\decibel}$ and $\epsilon_P = \SI{6}{\decibel}$.}
\label{fig:rates_mac}
\end{figure}

Fig.~\ref{fig:rates_mac} shows the rates achieved by the two users.
This figure also shows the capacity region and the \gls{OMA} capacity, both achieved by Gaussian modulation and an optimal non-\gls{BMD} receiver, making these rates unachievable by practical systems. For comparison, we also show the achievable rate of \gls{OMA} with 16\gls{QAM} and an \gls{BMD} receiver.
As one can see, the E2E system outperforms the rates enabled by the latter for the three considered values of $w$.
The highest \gls{ACLR} achieved by the E2E system is \SI{-29.95}{\decibel} and the \gls{CCDF} of the \gls{PAPR} at \SI{5.60}{\decibel} is $10^{-3}$, showing that the constraints have the expected impact on the learned waveforms.

\begin{figure}[h]
\centering
\begin{tikzpicture}

\definecolor{color0}{rgb}{0.12156862745098,0.466666666666667,0.705882352941177}
\definecolor{color1}{rgb}{1,0.498039215686275,0.0549019607843137}

\begin{axis}[
legend cell align={left},
legend style={
  fill opacity=0.8,
  draw opacity=1,
  text opacity=1,
  at={(0.97,0.6)},
  anchor=north east,
  draw=white!80!black
},
tick align=outside,
tick pos=left,
x grid style={white!69.0196078431373!black},
xmajorgrids,
xmin=-1.5, xmax=1.5,
xtick style={color=black},
y grid style={white!69.0196078431373!black},
ymajorgrids,
ymin=-1.5, ymax=1.5,
ytick style={color=black}
]
\addplot [semithick, color0, mark=*, mark size=3, mark options={solid}, only marks]
table {%
-0.702398240566254 1.37632215023041
-0.739641129970551 1.36602425575256
-0.162087589502335 0.157502919435501
-0.169473931193352 0.149451494216919
-0.51291686296463 0.954731583595276
-0.508668601512909 0.950898885726929
-0.316275537014008 0.50470232963562
-0.315900653600693 0.487536609172821
0.545136332511902 -1.44059026241302
0.558002591133118 -1.43959772586823
0.0296413395553827 -0.256267309188843
0.0242941472679377 -0.259843766689301
0.384819537401199 -1.02439069747925
0.380364924669266 -1.03578293323517
0.165637493133545 -0.588035047054291
0.171745240688324 -0.59892326593399
};
\addlegendentry{User 1}
\addplot [semithick, color1, mark=square*, mark size=3, mark options={solid}, only marks]
table {%
0.819900035858154 0.734968066215515
-1.01845014095306 -1.16956841945648
0.816279232501984 0.7347012758255
-1.01887595653534 -1.22100031375885
0.36645844578743 0.27254182100296
-0.203405737876892 -0.325950056314468
0.369900107383728 0.270727097988129
-0.191258504986763 -0.322805225849152
1.09413468837738 1.04368698596954
-0.752181112766266 -0.883357882499695
1.08043956756592 1.03715181350708
-0.732476830482483 -0.897709727287292
0.290097445249557 0.189334690570831
-0.278902947902679 -0.404301226139069
0.281028032302856 0.191186994314194
-0.287963300943375 -0.403320580720901
};
\addlegendentry{User 2}
\draw (axis cs:-0.702398240566254,1.37632215023041) node[
  scale=0.5,
  anchor=base west,
  text=black,
  rotate=0.0
]{0000};
\draw (axis cs:-0.739641129970551,1.36602425575256) node[
  scale=0.5,
  anchor=base west,
  text=black,
  rotate=0.0
]{0001};
\draw (axis cs:-0.162087589502335,0.157502919435501) node[
  scale=0.5,
  anchor=base west,
  text=black,
  rotate=0.0
]{0010};
\draw (axis cs:-0.169473931193352,0.149451494216919) node[
  scale=0.5,
  anchor=base west,
  text=black,
  rotate=0.0
]{0011};
\draw (axis cs:-0.51291686296463,0.954731583595276) node[
  scale=0.5,
  anchor=base west,
  text=black,
  rotate=0.0
]{0100};
\draw (axis cs:-0.508668601512909,0.950898885726929) node[
  scale=0.5,
  anchor=base west,
  text=black,
  rotate=0.0
]{0101};
\draw (axis cs:-0.316275537014008,0.50470232963562) node[
  scale=0.5,
  anchor=base west,
  text=black,
  rotate=0.0
]{0110};
\draw (axis cs:-0.315900653600693,0.487536609172821) node[
  scale=0.5,
  anchor=base west,
  text=black,
  rotate=0.0
]{0111};
\draw (axis cs:0.545136332511902,-1.44059026241302) node[
  scale=0.5,
  anchor=base west,
  text=black,
  rotate=0.0
]{1000};
\draw (axis cs:0.558002591133118,-1.43959772586823) node[
  scale=0.5,
  anchor=base west,
  text=black,
  rotate=0.0
]{1001};
\draw (axis cs:0.0296413395553827,-0.256267309188843) node[
  scale=0.5,
  anchor=base west,
  text=black,
  rotate=0.0
]{1010};
\draw (axis cs:0.0242941472679377,-0.259843766689301) node[
  scale=0.5,
  anchor=base west,
  text=black,
  rotate=0.0
]{1011};
\draw (axis cs:0.384819537401199,-1.02439069747925) node[
  scale=0.5,
  anchor=base west,
  text=black,
  rotate=0.0
]{1100};
\draw (axis cs:0.380364924669266,-1.03578293323517) node[
  scale=0.5,
  anchor=base west,
  text=black,
  rotate=0.0
]{1101};
\draw (axis cs:0.165637493133545,-0.588035047054291) node[
  scale=0.5,
  anchor=base west,
  text=black,
  rotate=0.0
]{1110};
\draw (axis cs:0.171745240688324,-0.59892326593399) node[
  scale=0.5,
  anchor=base west,
  text=black,
  rotate=0.0
]{1111};
\draw (axis cs:0.819900035858154,0.734968066215515) node[
  scale=0.5,
  anchor=base west,
  text=black,
  rotate=0.0
]{0000};
\draw (axis cs:-1.01845014095306,-1.16956841945648) node[
  scale=0.5,
  anchor=base west,
  text=black,
  rotate=0.0
]{0001};
\draw (axis cs:0.816279232501984,0.7347012758255) node[
  scale=0.5,
  anchor=base west,
  text=black,
  rotate=0.0
]{0010};
\draw (axis cs:-1.01887595653534,-1.22100031375885) node[
  scale=0.5,
  anchor=base west,
  text=black,
  rotate=0.0
]{0011};
\draw (axis cs:0.36645844578743,0.27254182100296) node[
  scale=0.5,
  anchor=base west,
  text=black,
  rotate=0.0
]{0100};
\draw (axis cs:-0.203405737876892,-0.325950056314468) node[
  scale=0.5,
  anchor=base west,
  text=black,
  rotate=0.0
]{0101};
\draw (axis cs:0.369900107383728,0.270727097988129) node[
  scale=0.5,
  anchor=base west,
  text=black,
  rotate=0.0
]{0110};
\draw (axis cs:-0.191258504986763,-0.322805225849152) node[
  scale=0.5,
  anchor=base west,
  text=black,
  rotate=0.0
]{0111};
\draw (axis cs:1.09413468837738,1.04368698596954) node[
  scale=0.5,
  anchor=base west,
  text=black,
  rotate=0.0
]{1000};
\draw (axis cs:-0.752181112766266,-0.883357882499695) node[
  scale=0.5,
  anchor=base west,
  text=black,
  rotate=0.0
]{1001};
\draw (axis cs:1.08043956756592,1.03715181350708) node[
  scale=0.5,
  anchor=base west,
  text=black,
  rotate=0.0
]{1010};
\draw (axis cs:-0.732476830482483,-0.897709727287292) node[
  scale=0.5,
  anchor=base west,
  text=black,
  rotate=0.0
]{1011};
\draw (axis cs:0.290097445249557,0.189334690570831) node[
  scale=0.5,
  anchor=base west,
  text=black,
  rotate=0.0
]{1100};
\draw (axis cs:-0.278902947902679,-0.404301226139069) node[
  scale=0.5,
  anchor=base west,
  text=black,
  rotate=0.0
]{1101};
\draw (axis cs:0.281028032302856,0.191186994314194) node[
  scale=0.5,
  anchor=base west,
  text=black,
  rotate=0.0
]{1110};
\draw (axis cs:-0.287963300943375,-0.403320580720901) node[
  scale=0.5,
  anchor=base west,
  text=black,
  rotate=0.0
]{1111};
\end{axis}

\end{tikzpicture}
\caption{Learned constellations for two interfering users ($w = 0.5$, $\epsilon_A = \SI{-30}{\decibel}$, $\epsilon_P = \SI{6}{\decibel}$).}
\label{fig:mac_const}
\end{figure}

\begin{figure}[h]
\centering{
\input{figs/mac_filters.tex}
\caption{The two transmit filters and single receive filter learned over the interference channel, with $w = 0.5$, $\epsilon_A = \SI{-30}{\decibel}$, and $\epsilon_P = \SI{6}{\decibel}$.}
\label{fig:mac_filters}
}
\end{figure}

To better understand how the E2E system operates, we show in Fig.~\ref{fig:mac_const} and Fig.~\ref{fig:mac_filters}, respectively, the learned constellations and filters for $w = 0.5$.
As one can see, the constellations are along close-to orthogonal axes.
Moreover, the learned transmit filters are also only partially overlapping, as each user favors a distinct part of the available band. An in-depth study of the learned filters and constellations is out of scope of this paper and left for future investigation.


\subsection{Multipath channel}
\label{sec:mp}

The use of multi-carrier waveforms, such as \gls{OFDM}, for multipath channels  has been mainly motivated by the low-complexity single-tap equalization they enable.
However, with the advent of neural receivers that can be efficiently implemented on dedicated hardware~\cite{10.1145/3361682}, this advantage might become less relevant, especially as multi-carrier waveforms are well-known to incur high \glspl{PAPR}, leading to power amplifier inefficiencies.

In order to illustrate this, we consider a multipath channel in this section.
We simulate the channel according to the \gls{3GPP} 38.901 UMi \gls{LOS} and \gls{NLOS} models, based on a dataset of baseband channel coefficients $a_p$ and corresponding delays $\tau_p$ which were generated with the help of the Quadriga channel simulator~\cite{quadriga}.
The channel is assumed to be static for the duration of a block, corresponding to a low-mobility scenario.
Training and benchmarking were performed on two different datasets to ensure that the E2E system is able to generalize to previously unseen channel realizations.
Although over-the-air training using the technique described in \cite{aoudia2019model} might be possible, we believe that our approach is most practical if training is done offline, prior to deployment. 

An \gls{OFDM} waveform with Gray-labeled 16\gls{QAM}, \gls{LMMSE} channel estimation, and single-tap equalization was implemented to serve as BL, with a subcarrier spacing of \SI{30}{\kilo\Hz} and 166~subcarriers.
Each block consists of one slot, i.e., 14 \gls{OFDM} symbols.
A standard \gls{5GNR} pilot pattern was used, with an overhead of $1/14$ ($\approx$ \SI{7.14}{\percent}).

For the E2E system, the neural receiver from Fig.~\ref{fig:nnrx} was used.
The first $N_P = 32$ samples of each block were used as trainable pilots, which were centered and normalized as for the constellation.
The length of the vector $\zv$ in Fig.~\ref{fig:nnrx}, which is generated by the dense network from the received pilot symbols, was set to $N_S = 64$.
Training was done over the range of \glspl{SNR} from \SI{5}{\decibel} to \SI{20}{\decibel}, i.e., the \gls{SNR} was randomly sampled for each training example.

\begin{figure}[h]
\centering{
\begin{tikzpicture}

\definecolor{color0}{rgb}{0.12156862745098,0.466666666666667,0.705882352941177}
\definecolor{color1}{rgb}{1,0.498039215686275,0.0549019607843137}
\definecolor{color2}{rgb}{0.172549019607843,0.627450980392157,0.172549019607843}

\begin{axis}[
legend cell align={left},
legend style={
  fill opacity=0.8,
  draw opacity=1,
  text opacity=1,
  at={(0.15,0.01)},
  anchor=south west,
  draw=white!80!black
},
tick align=outside,
tick pos=left,
x grid style={white!69.0196078431373!black},
xlabel={SNR [\si{\decibel}]},
xmajorgrids,
xmin=4.28260869565217, xmax=20.0652173913044,
xtick style={color=black},
y grid style={white!69.0196078431373!black},
ylabel={\(\displaystyle R\) [\si{\bit\per\second\per\hertz}]},
ymajorgrids,
ymin=1.34261627793312, ymax=3.54102842211723,
ytick style={color=black}
]
\addplot [semithick, color0, mark=*, mark size=3, mark options={solid}]
table {%
5 1.44864937887911
6.30434782608696 1.70743849553036
7.60869565217391 1.95407826158055
8.91304347826087 2.18441215181427
10.2173913043478 2.44399450162058
11.5217391304348 2.65125711676014
12.8260869565217 2.87103803295576
14.1304347826087 3.02330741625919
15.4347826086957 3.13994345312899
16.7391304347826 3.22345983790739
18.0434782608696 3.27888168817921
19.3478260869565 3.33125696155462
};
\addlegendentry{OFDM $\text{\gls{ACLR}} = \SI{-22.04}{\decibel}$}
\addplot [semithick, color1, mark=diamond*, mark size=3, mark options={solid}]
table {%
5 1.47799754142761
6.30434782608696 1.69814515113831
7.60869565217391 1.9953361749649
8.91304347826087 2.21504092216492
10.2173913043478 2.50704264640808
11.5217391304348 2.70255875587463
12.8260869565217 2.94874739646912
14.1304347826087 3.08680939674377
15.4347826086957 3.2535240650177
16.7391304347826 3.32231688499451
18.0434782608696 3.41595554351807
19.3478260869565 3.44110059738159
};
\addlegendentry{E2E $\epsilon_A = \SI{-20}{\decibel}$, $\text{\gls{ACLR}} = \SI{-20.07}{\decibel}$}
\addplot [semithick, color2, mark=square*, mark size=3, mark options={solid}]
table {%
5 1.44254410266876
6.30434782608696 1.65859115123749
7.60869565217391 1.94953072071075
8.91304347826087 2.16680002212524
10.2173913043478 2.46110820770264
11.5217391304348 2.65922117233276
12.8260869565217 2.91367650032043
14.1304347826087 3.0544159412384
15.4347826086957 3.23345112800598
16.7391304347826 3.30268788337708
18.0434782608696 3.40630102157593
19.3478260869565 3.42988348007202
};
\addlegendentry{E2E $\epsilon_A =\SI{-30}{\decibel}$, $\text{\gls{ACLR}} = \SI{-30.09}{\decibel}$}
\end{axis}

\end{tikzpicture}
\caption{Rates achieved by the \gls{OFDM} BL and the two E2E systems trained with different \gls{ACLR} constraints and $\epsilon_P = \SI{7}{\decibel}$. The \gls{ACLR} achieved by each waveform is indicated in the legend.}
\label{fig:rates_mp}
}
\end{figure}

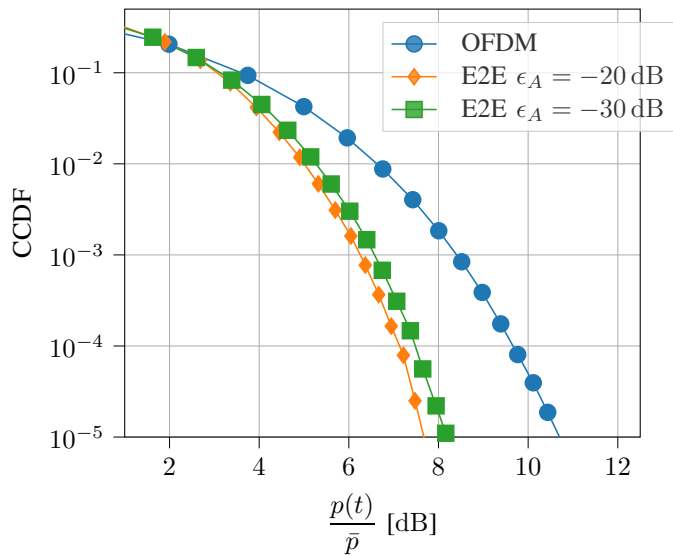
\begin{figure}[h]
\centering{
\begin{tikzpicture}

\definecolor{color0}{rgb}{0.12156862745098,0.466666666666667,0.705882352941177}
\definecolor{color1}{rgb}{1,0.498039215686275,0.0549019607843137}
\definecolor{color2}{rgb}{0.172549019607843,0.627450980392157,0.172549019607843}

\begin{axis}[
legend cell align={left},
legend style={
  fill opacity=0.8,
  draw opacity=1,
  text opacity=1,
  at={(0.5,0.97)},
  anchor=north west,
  draw=white!80!black
},
log basis y={10},
tick align=outside,
tick pos=left,
x grid style={white!69.0196078431373!black},
xlabel={\(\displaystyle \frac{p(t)}{\bar{p}}\) [\si{\decibel}]},
xmajorgrids,
xmin=1, xmax=12.5,
xtick style={color=black},
y grid style={white!69.0196078431373!black},
ylabel={CCDF},
ymajorgrids,
ymin=1e-05, ymax=0.5,
ymode=log,
ytick style={color=black},
ytick={1e-06,1e-05,0.0001,0.001,0.01,0.1,1,10},
yticklabels={
  \(\displaystyle {10^{-6}}\),
  \(\displaystyle {10^{-5}}\),
  \(\displaystyle {10^{-4}}\),
  \(\displaystyle {10^{-3}}\),
  \(\displaystyle {10^{-2}}\),
  \(\displaystyle {10^{-1}}\),
  \(\displaystyle {10^{0}}\),
  \(\displaystyle {10^{1}}\)
}
]
\addplot [semithick, color0, mark=*, mark size=3, mark options={solid}]
table {%
-141.110275268555 0.999999942673699
-1.02477931976318 0.454075384086219
1.98552072048187 0.206096824122908
3.7464337348938 0.0936143086448062
4.99581956863403 0.0424790185737216
5.96492004394531 0.0192831919284567
6.7567343711853 0.00878227470763582
7.42621040344238 0.00403663150653522
8.00613689422607 0.00184074753496899
8.5176477432251 0.000841034166475629
8.97529220581055 0.000387239165329012
9.38923072814941 0.000174845218986475
9.7673397064209 8.04861270350399e-05
10.1149482727051 3.94404952992478e-05
10.4367017745972 1.8688374226139e-05
10.7366237640381 9.17220820917386e-06
11.0168209075928 3.89818848889334e-06
11.3152265548706 1.71978903917847e-06
11.5348815917969 6.87915615693591e-07
11.7684288024902 4.58610410425386e-07
12.3488521575928 1.71978903895642e-07
12.3733444213867 1.14652602634102e-07
12.4296360015869 5.73263012615399e-08
12.44309425354 0
};
\addlegendentry{OFDM}
\addplot [semithick, color1, mark=diamond*, mark size=3, mark options={solid}]
table {%
-61.2136421203613 0.999999
-5.09430885314941 0.787886
-2.08406734466553 0.635395
-0.323164105415344 0.478457
0.926233768463135 0.326808
1.89531350135803 0.218393
2.68712472915649 0.137294
3.35659456253052 0.077709
3.93653345108032 0.041637
4.44810485839844 0.022217
4.90568208694458 0.011783
5.31959819793701 0.00603100000000001
5.69749355316162 0.00310600000000005
6.0451979637146 0.001614
6.36752557754517 0.000773000000000024
6.6672945022583 0.000365999999999977
6.94697856903076 0.000164999999999971
7.21766757965088 7.90000000000513e-05
7.47639942169189 2.50000000000528e-05
7.72349452972412 8.000000000008e-06
8.06445503234863 1.99999999994649e-06
8.12966537475586 1.00000000002876e-06
8.37348556518555 0
8.37348556518555 0
};
\addlegendentry{E2E $\epsilon_A = \SI{-20}{\decibel}$}
\addplot [semithick, color2, mark=square*, mark size=3, mark options={solid}]
table {%
-76.9208221435547 0.999999
-4.39327669143677 0.758828
-1.38308465480804 0.561191
0.37783071398735 0.385452
1.62721109390259 0.245732
2.59631824493408 0.147065
3.38814258575439 0.082999
4.05759429931641 0.044848
4.63751077651978 0.023354
5.14907789230347 0.011924
5.60672569274902 0.00600500000000004
6.02072620391846 0.00302000000000002
6.39986085891724 0.00146800000000002
6.74613475799561 0.000677999999999956
7.06865549087524 0.000310000000000032
7.37236309051514 0.000147000000000008
7.64864444732666 5.5999999999945e-05
7.94694852828979 2.19999999999665e-05
8.16226577758789 1.09999999999832e-05
8.44120597839355 5.00000000003276e-06
8.69479560852051 1.99999999994649e-06
9.07448482513428 0
9.07448482513428 0
};
\addlegendentry{E2E $\epsilon_A = \SI{-30}{\decibel}$}
\end{axis}

\end{tikzpicture}
\caption{Power distribution of \gls{OFDM} and two E2E systems trained with different \gls{ACLR} constraints and $\epsilon_P = \SI{7}{\decibel}$.}
\label{fig:ccdf_mp}
}
\end{figure}

Fig.~\ref{fig:rates_mp} shows the rates achieved by the BL and the E2E system for $\epsilon_P$ set to \SI{7}{\decibel} and $\epsilon_A$ to \SI{-20}{\decibel} and \SI{-30}{\decibel}.
The \gls{ACLR} achieved by each waveform is indicated in the legend.
The E2E system achieves rates competitive or higher than the ones of the BL, especially at high \glspl{SNR}.
This is particularly interesting for $\epsilon_A =\SI{-30}{\decibel}$, where the E2E system enables an \gls{ACLR} that is \SI{10}{\decibel} lower than that of the BL.
Regarding the \gls{PAPR}, Fig.~\ref{fig:ccdf_mp} shows the \gls{CCDF} of the normalized power distribution of the compared waveforms.
As one can see, the E2E system achieves a \SI{2}{\decibel} gain over \gls{OFDM} for a \gls{CCDF} value of $10^{-4}$.

\begin{figure*}[h]
\centering{
\def\svgwidth{0.8\linewidth}
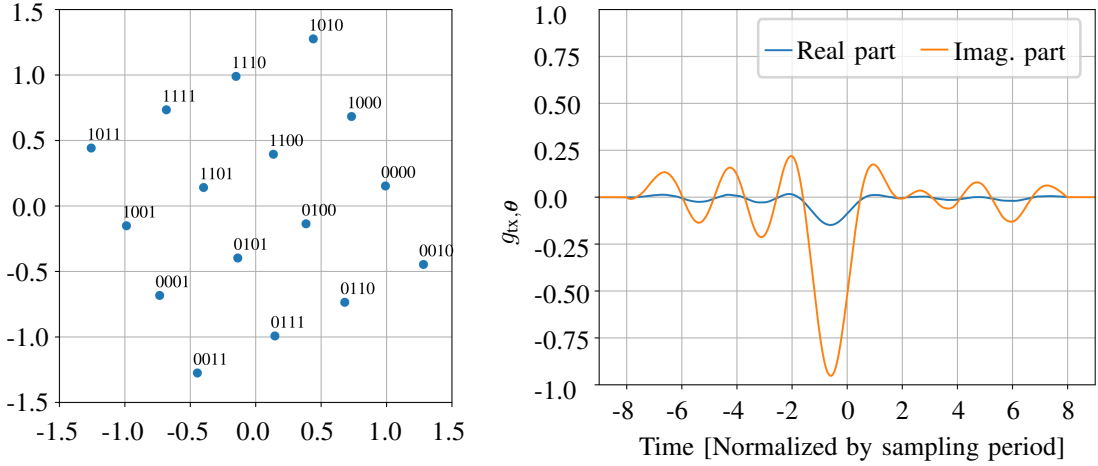
\caption{Learned constellation, corresponding bit labeling, and transmit filter for the multipath channel with $\epsilon_A = \SI{-30}{\decibel}$ and $\epsilon_P = \SI{7}{\decibel}$.}
\label{fig:mp_const_tmp}
}
\end{figure*}

\raggedbottom

Fig.~\ref{fig:mp_const_tmp} shows the learned constellation, corresponding labeling, and transmit filter for $\epsilon_A = \SI{-30}{\decibel}$ and $\epsilon_P = \SI{7}{\decibel}$.
One can see that the E2E system has learned a Gray labeled 16\gls{QAM} constellation.
Interpreting the learned transmit filter is more difficult and left to future work. One can notice that it has similarities with a conventional \gls{RRC} filter, e.g., the presence of a central peak and adjacent ripples, but differs through the presence of a real and imaginary component and the lack of symmetry around the peak.

\section{Conclusion}
\label{sec:cl}

We have introduced a new method for designing single-carrier waveforms for communication systems through joint optimization of the transmit filter, receive filter, constellation geometry, corresponding bit labeling, and a neural receiver.
The training process was formulated as an optimization problem with constraints on the \gls{ACLR} and \gls{PAPR}, where the objective function is an achievable rate of practical systems using \gls{BMD} receivers.
Our results show that the proposed method enables fine control of the tradeoff between the information rate, \gls{ACLR}, and \gls{PAPR}.
On the \gls{AWGN} channel, the learned waveforms achieve rates that are competitive with the ones of conventional waveforms, but with significantly lower \glspl{ACLR}.
Lower \glspl{PAPR} are also possible, however at the cost of a rate loss.
We have further illustrated how the proposed method could enable a new approach for multiplexing of multiple users through the joint optimization of constellation geometries and transmit filters.
We have also shown that the learned single-carrier waveforms are competitive with \gls{OFDM} for multipath channels. They enable similar or higher rates, but with significantly lower \glspl{PAPR} and \glspl{ACLR}.
The interpretation of the learned filters reminds an open problem that requires future investigation.
Key directions of future investigation include the extension to other channel models (e.g., optical, sub-THz, MIMO), new use cases (e.g., faster-than-Nyquist signaling), as well as prototyping with software-defined radios.

\section*{Appendix}
\label{sec:app}

\subsection{Derivation of~\eqref{eq:avg_sig_pow}}

\textcolor{black}{
The signal power at time $t$ is
\begin{align}
	\bar{p}(t) \coloneqq \EE \LSB \abs{x(t)}^2 \RSB &= \EE \LSB \abs{\sum_{n=0}^{N-1} s_n g_{\text{tx},\thetav}(t-nT)}^2 \RSB\\
			   &= \sum_{n=0}^{N-1} \abs{g_{\text{tx},\thetav}(t-nT)}^2
\end{align}
where the second equality comes from the definition of $x(t)$~\eqref{eq:x} and the third equality follows from the baseband symbols $s_n$ being \gls{iid} with unit variance.
The total duration of the transmitted signal $x(t)$ is $(N-1)T + D$, where $T$ is the sampling period.
Without loss of generality, it is assumed that the signal starts at $t=0$.
The avarage signal power is therefore
\begin{align}
	\bar{p} &= \frac{\int_0^{(N-1)T + D}\bar{p}(t)dt}{(N-1)T + D}\\
			&= \frac{N}{(N-1)T + D}
\end{align}
where the second equality comes from the transmit filter having unit energy $\int_{-\frac{D}{2}}^{\frac{D}{2}} \abs{g_{\text{tx},\thetav}(t)}^2 dt = 1$.}

\subsection{Derivation of~\eqref{eq:c_norm}}

The transmit filter is described in the time domain as
\begin{equation}
	g_{\text{tx},\thetav}(t) = \frac{\sqrt{C(\thetav)}}{D} \text{rect}\left( \frac{t}{D} \right) \sum_{s=-S}^S \theta_s e^{j2\pi\frac{s}{D}t}.
\end{equation}
Its total energy can be computed as
\begin{align}
E_{T}(\thetav) &= \int_{-\frac{D}{2}}^{\frac{D}{2}} \abs{g_{\text{tx},\thetav}(t)}^2 dt\\
&= \frac{C(\thetav)}{D^2} \sum_{s_1=-S}^S \sum_{s_2=-S}^S \theta_{s_1} \theta_{s_2}^* \int_{-\frac{D}{2}}^{\frac{D}{2}} e^{j\frac{2\pi}{D}(s_1 - s_2)t} dt.
\end{align}
The integral of the last equation is
\begin{equation}
\int_{-\frac{D}{2}}^{\frac{D}{2}} e^{j\frac{2\pi}{D}(s_1-s_2)t} dt =
\begin{cases}
	0,& \text{ if } s_1 \neq s_2\\
	D,& \text{ if } s_1 = s_2
\end{cases}
\end{equation}
which leads to
\begin{equation}
E_{T}(\thetav) = \frac{C(\thetav)}{D} \thetav\htp \thetav.
\end{equation}
To ensure that the transmit filter has unit energy, the normalization constant is hence set to
\begin{equation}
	C(\thetav) = \frac{D}{\thetav\htp \thetav}.
\end{equation}

\subsection{Derivation of~\eqref{eq:tx_rx_conv}}

One has
\begin{multline}
	\LB g_{\text{tx},\thetav} \ast g_{\text{rx},\psiv}^* \RB (t) = \int g_{\text{tx},\thetav}(w) g_{\text{rx},\psiv}^* \LB t - w \RB dw\\
	= \frac{\sqrt{C(\thetav)}}{D^2}\sum_{s_1=-S}^S \sum_{s_2=-S}^S \psi_{s_1}^* \theta_{s_2}\cdot\\
	\int_{-\frac{D}{2}}^{\frac{D}{2}} \text{rect}\LB \frac{t-w}{D}\right) e^{j\frac{2\pi}{D}\left( s_2 w - (t-w)s_1 \RB} dw
\end{multline}
which equals 0 if $t \notin \LB -D, D \RB$.
Assuming $t \in \LB -D, D \RB$, let us define $z \coloneqq \frac{t-w}{D}$. Then
\begin{multline}
\LB g_{\text{tx},\thetav} \ast g_{\text{rx},\psiv}^* \RB (t) = \frac{\sqrt{C(\thetav)}}{D}\sum_{s_1=-S}^S \sum_{s_2=-S}^S \psi_{s_1}^* \theta_{s_2} e^{j2\pi\frac{s_2}{D}t}\cdot\\
\int_{\frac{t}{D}-\frac{1}{2}}^{\frac{t}{D} + \frac{1}{2}} \text{rect}(z) e^{-j2\pi(s_1+s_2)z} dz.
\end{multline}
By introducing $I_{min}(t) \coloneqq \text{max}\left(-\frac{1}{2};\frac{t}{D}-\frac{1}{2}\right)$ and $I_{max}(t) \coloneqq \text{min}\left(\frac{1}{2};\frac{t}{D}+\frac{1}{2}\right)$, one has
\begin{multline}
\LB g_{\text{tx},\thetav} \ast g_{\text{rx},\psiv}^* \RB (t) = \frac{\sqrt{C(\thetav)}}{D}\sum_{s_1=-S}^S \sum_{s_2=-S}^S \psi_{s_1}^* \theta_{s_2} e^{j2\pi\frac{s_2}{D}t}\cdot\\
	\int_{I_{min}(t)}^{I_{max}(t)}e^{-j2\pi(s_1+s_2)z} dz\\
	= \frac{\sqrt{C(\thetav)}}{D} \thetav\tp \Am(t) \psiv
\end{multline}
where $\Am(t)$ is a $(2S+1)\times(2S+1)$ complex-valued matrix such that
\begin{multline}
	A(t)_{s_1,s_2} = e^{j2\pi\frac{s_2}{D}t} \int_{I_{min}(t)}^{I_{max}(t)}e^{-j2\pi(s_1+s_2)z} dz\\
	=\begin{cases}
	e^{j2\pi\frac{s_2}{D}t}\Delta(t) &\text{ if } s_1 + s_2 = 0\\
	e^{j\pi\LB 2\frac{s_2}{D}t - (s_1+s_2)\Sc(t)\RB}\frac{\sin \LB \pi(s_1+s_2)\Delta(t) \RB}{\pi(s_1+s_2)} &\text{ otherwise}
	\end{cases}
\end{multline}
and $\Delta(t) \coloneqq I_{max}(t) - I_{min}(t)$, $\Sc(t) \coloneqq I_{max}(t) + I_{min}(t)$.

\subsection{Derivation of~\eqref{eq:noise_conv_filt}}
The derivation follows the same steps as that of \eqref{eq:tx_rx_conv} and is not shown for brevity.

\subsection{Derivation of~\eqref{eq:ei_trf}}\label{sec:derivation_aclr}

The in-band energy of the transmitted signal is
\begin{multline}
	E_{I}(\thetav) = \int_{-\frac{W}{2}}^{\frac{W}{2}} \abs{\hat{g}_{\text{tx},\thetav}(f)}^2 df\\
	= C(\thetav) \sum_{s_1=-S}^{S} \sum_{s_2=-S}^S \theta_{s_1}^* \theta_{s_2}\cdot\\
	\int_{-\frac{W}{2}}^{\frac{W}{2}} \text{sinc}(Df-s_1) \text{sinc}(Df-s_2)df\\
	= C(\thetav) \thetav\htp \Em \thetav
\end{multline}
where $\Em$ is the real-valued $(2S+1) \times (2S+1)$ matrix with elements
\begin{equation}
	E_{s_1,s_2} = \int_{-\frac{W}{2}}^{\frac{W}{2}} \text{sinc}(Df-s_1) \text{sinc}(Df-s_2)df
\end{equation}
which can be easily pre-computed for a given bandwidth $W$ and filter duration $D$.

\bibliographystyle{IEEEtran}
\bibliography{IEEEabrv,bibliography}

\end{document}